\documentclass[a4paper,11pt]{article}

\usepackage{jheppub}
\usepackage{amsfonts}
\usepackage{dsfont}
\usepackage{amsmath,amssymb}

\usepackage{bbm}
\usepackage{bm}
\usepackage{color}
\usepackage{slashed}
\usepackage{siunitx}
\DeclareSIUnit \kpc {kpc}
\usepackage{booktabs}
\usepackage{physics}
\usepackage{array}
\usepackage{subcaption}
\usepackage[utf8]{inputenc}
\usepackage{graphicx}
\usepackage{bbold}
\usepackage{bm}
\usepackage{todonotes}
\usepackage{enumitem}
\usepackage[normalem]{ulem}
\DeclareGraphicsExtensions{.pdf,.png,.jpg}

\newcommand{\SBcomment}[1]{\emph{\color{blue}SB:#1}}

\newcommand{\rmi}[1]{{\mbox{\scriptsize #1}}}
\newcommand{\rmii}[1]{{\mbox{\tiny\rm{#1}}}}
\newcommand{\mDM}{m_{\rmii{DM}}}
\newcommand{\muDM}{\mu_{\rmii{DM}}}

\newcommand{\vrel}{v_\rmi{rel}}

\title{Dark matter bound states made easy}

\author[a]{Simone Biondini,}
\author[a]{and Stefan Vogl}

\affiliation[a]{Institute of Physics, University of Freiburg, Hermann-Herder-Straße 3, 79014 Freiburg, Germany}

\emailAdd{simone.biondini@physik.uni-freiburg.de}
\emailAdd{stefan.vogl@physik.uni-freiburg.de}

\abstract{
A light force carrier can mediate long-range interactions between dark matter particles and facilitate the formation of bound states. Models incorporating these features have attracted considerable attention in recent years, with significant advances in both the theoretical understanding of such bound states and their impact on the parameter space consistent with the observed dark matter relic abundance. However, the relevant results and their consistent implementation in Boltzmann equations remain technically intricate, requiring careful treatment or the use of specialized tools. This complexity has hindered broader adoption of these findings  and restricted their use to a small number of expert groups. To address this challenge, we propose a simple prescription based on the analytic limits of the relevant rates, which allows standard freeze-out computations to be refined through a straightforward semi-analytical prescription that captures the main features of the full calculation. We illustrate this approach for a $U(1)$ dark matter model with a light force carrier, including Sommerfeld enhancement and bound-state effects.
}

\begin{document}
\maketitle
\flushbottom

\section{Introduction}
\label{sec:intro}

The determination of the dark matter (DM) relic abundance from thermal freeze-out remains one of the most important links between particle physics and cosmology. In its simplest form, the calculation relies on perturbative annihilation cross sections and leads to the well-known weakly interacting massive particle (WIMP) paradigm. The central quantity that enters the evolution equations of the dark matter number density, and ultimately enables a prediction for the dark matter cosmological abundance,
is the thermally averaged annihilation cross section \cite{Gondolo:1990dk,Griest:1990kh}. 

However, it has become increasingly clear that special care is required in scenarios where dark matter interacts through light mediators. In such cases, non-perturbative effects near threshold can substantially modify the freeze-out dynamics and significantly affect the relic density, with corrections that can easily exceed an order of magnitude.
Two classes of effects are particularly relevant. First, long-range interactions distort the wave function of the incoming particles, leading to the Sommerfeld enhancement of annihilation cross sections~\cite{Sommerfeld:1931qaf,Hisano:2004ds} for an attractive potential. Second, sufficiently strong attractive forces allow for the formation of metastable bound states, whose subsequent decay provides an additional depletion channel for dark matter~\cite{Feng:2009mn,vonHarling:2014kha}.
Both effects originate from the same underlying non-relativistic dynamics and can naturally be described within non-relativistic effective field theories (NREFTs) \cite{Bodwin:1994jh,Pineda:1997bj,Brambilla:1999xf,Brambilla:2004jw}. 

Over the past decade, significant theoretical progress has been made in understanding and systematically including these near-threshold effects. Early studies established the importance of bound-state formation for the determination of the dark matter relic density in scenarios with abelian vector mediators \cite{Feng:2009mn,vonHarling:2014kha,Petraki:2016cnz,Binder:2019erp,Biondini:2023zcz}, and this work was later extended to non-abelian gauge groups \cite{Mitridate:2017izz,Harz:2018csl,Liew:2016hqo,Ellis:2015vaa,Ellis:2015vna,Biondini:2018pwp,Biondini:2023zcz,Garny:2021qsr} and scalar mediators \cite{Oncala:2018bvl,Oncala:2019yvj,Harz:2018csl,Biondini:2018xor,Biondini:2021ycj,Biondini:2021ccr,Oncala:2021swy}. The impact of bound-state formation on dark matter phenomenology, as well as its interplay with experimental constraints, has been investigated across a variety of dark matter models \cite{Mitridate:2017izz,Biondini:2018ovz,Ko:2019wxq,Biondini:2023yxt,Biondini:2019int,Garny:2021qsr,Biondini:2023ksj,Baumgart:2023pwn,Bollig:2021psb,Biondini:2025gpg,Biondini:2026ryb}.\footnote{An example of direct-detection phenomenology for stable DM bound states can be found in ref.~\cite{Laha:2013gva}.} More recently, a systematic treatment has been developed based on NREFTs and their thermal extensions \cite{Escobedo:2008sy,Escobedo:2010tu,Brambilla:2008cx,Biondini:2018ovz,Binder:2020efn}. In particular, bound-state formation induced by scattering with plasma constituents and various thermal effects, including thermal Debye masses, have been addressed in Refs.~\cite{Binder:2020efn,Biondini:2025jvp,Biondini:2024aan,Biondini:2018pwp,Binder:2018znk}, while careful extractions of the potential between DM pairs and the associated Sommerfeld factors can be found in Refs.~\cite{Beneke:2012tg,Beneke:2014gja,Blum:2016nrz,Flores:2024sfy,Bhattacharya:2026sfe}. Finally, bound-to-bound transitions have been studied in detail in Refs.~\cite{Garny:2021qsr,Binder:2021vfo,Binder:2023ckj,Biondini:2025jvp}. Altogether, these developments have led to a significantly more refined and quantitatively reliable description of dark matter freeze-out in the presence of long-range interactions. 

At the same time, there has been important progress in bridging these theoretical developments with practical tools for phenomenological studies. Recent efforts have explored the implementation of bound-state effects in numerical frameworks and public codes~\cite{Becker:2026icc,Becker:2025vgq,Binder:2025daq}, including approaches that interface with widely used tools such as \textrm{micrOMEGAs} \cite{Belanger:2006is,Belanger:2001fz}. These advances represent an important step toward making precision calculations of the relic density more broadly accessible.

Despite this progress, the practical implementation of near-threshold effects remains technically involved. The computation of the relevant rates, including Sommerfeld-enhanced annihilation, bound-state formation and dissociation, and transitions among bound states, requires a careful treatment of the non-relativistic dynamics of interacting particle pairs in a thermal environment. This typically entails handling multiple physical scales and their interplay, often leading to computationally demanding numerical evaluations. As a consequence, the use of these results is still largely confined to specialized studies, and their broader adoption in phenomenological analyses remains limited. At the same time, it is highly desirable to have simple and efficient ways to estimate the impact of near-threshold effects in a given model.

The goal of this work is to fill this gap and complement existing approaches by providing a simplified and transparent framework for incorporating bound-state effects in dark matter freeze-out. Rather than aiming at further increasing the level of theoretical sophistication, we focus on identifying the dominant physical contributions and developing analytic approximations that capture the essential features of the dynamics with minimal computational effort. While this approach necessarily involves a loss of precision, we carefully assess the associated uncertainties and quantify the accuracy of our approximations.

Our approach combines three main ingredients. First, we employ the potential NREFT (pNREFT) framework to organize the relevant processes and clarify their parametric behavior.\footnote{Even though the use of pNREFT is convenient here, these results can also be obtained with traditional methods, see e.g. \cite{Sommerfeld:1931qaf} for the original derivation of the Sommerfeld effect or \cite{Bethe:1957ncq} for an extensive discussion of bound state effects.}
Second, we construct simple analytic approximations for the key rates, in particular for bound-state dissociation, which allow us to avoid numerically expensive integrations while maintaining good accuracy. Third, we combine this with a semi-analytic treatment of the Boltzmann equation that enables both an efficient evaluation of the relic density and can be readily interfaced with existing freeze-out calculations. In particular, this approach reduces the determination of the dark matter yield to the numerical evaluation of a single integral, rather than requiring the solution of a differential Boltzmann equation.

The outcome is a practical strategy that allows one to include Sommerfeld and bound-state effects in a controlled and user-friendly way, without the need to solve the full system of coupled Boltzmann equations or to rely on specialized numerical implementations. In this sense, our work is intended as a complementary tool to existing approaches: it provides a simple “recipe” for practitioners that captures the dominant physics across a wide range of temperatures and couplings, while retaining good accuracy in the predicted relic abundance.

The paper is organized as follows. In section~\ref{sec:rates}, we summarize the relevant interaction rates within the pNREFT framework. In section~\ref{sec:effective_cross_approximate_rates}, we discuss the effective description of bound states and introduce simplified expressions for the key quantities, including semi-analytic approximations. Section~\ref{sec:numerical} explores the numerical solution of the Boltzmann equations with exact and approximated effective cross sections and  their impact on the relic density. In sections~\ref{sec:strategy_one} and \ref{sec:strategy_two}, we compare numerical and semi-analytical solutions of the Boltzmann equation and assess the accuracy of our approach. We conclude in section~\ref{sec:conclusions}.
\section{Summary of the particle interaction rates}
\label{sec:rates}
In order to set up our framework, we need a concrete particle physics model for dark matter. We focus on an abelian dark matter model, in line with much of the original and more recent literature on the subject. More specifically, the dark matter candidate is a Dirac fermion $X$ with mass $\mDM$, charged under an abelian gauge group U(1)$_{\rmii{DM}}$. For simplicity, we denote the corresponding dark photon by $\gamma$~\cite{Feldman:2006wd,Fayet:2007ua,Goodsell:2009xc,Morrissey:2009ur,Andreas:2011in}. Such a setup is widely used to investigate near-threshold effects in a simple yet sufficiently rich framework \cite{vonHarling:2014kha,Binder:2018znk,Binder:2020efn,Biondini:2025jvp}.

In addition, we allow for $n_f$ light Dirac fermion species $f_i$ as part of the dark sector field content, with $i=0,\ldots,n_f$, also charged under U(1)$_{\rmii{DM}}$. The case $n_f=0$ corresponds to the minimal abelian dark matter and dark photon scenario. These fermions are assumed to be relativistic, i.e.\ $m_i \ll T$, during freeze-out and the subsequent thermal evolution. Together with the dark photons, they constitute a radiation bath for the dark sector. 
This particle content allows us to systematically account not only for bound-state formation via gauge boson emission \cite{vonHarling:2014kha,Petraki:2015hla}, but also for processes induced by scatterings with light particles in the thermal bath \cite{Biondini:2017ufr,Biondini:2018pwp,Binder:2019erp,Binder:2020efn}.

The Lagrangian density is given by
\begin{equation}
\mathcal{L}=\bar{X} (i \slashed {D} -\mDM) X -\frac{1}{4} F_{\mu \nu} F^{\mu \nu} + \sum_{i=0}^{n_f}\bar{f}_i(i \slashed{D}-m_i) f_i + \mathcal{L}_{\textrm{portal}} \, ,
\label{lag_mod_0}
\end{equation}
where the covariant derivative reads $D_\mu=\partial_\mu + i g A_\mu$, with $A_\mu$ the dark photon field and $F_{\mu \nu} = \partial_\mu A_\nu - \partial_\nu A_\mu$. 
The term $\mathcal{L}_{\textrm{portal}}$ contains the portal(s) to the SM sector which we leave unspecified here.
We define the fine-structure constant as $\alpha \equiv g^2/(4 \pi)$ and work in the perturbative regime $\alpha \ll 1$. In our numerical analysis, we will set $m_i=0$ for all light fermion species and restrict ourselves to $\alpha \leq 0.2$.

Our main focus is to describe the dynamics of dark matter pairs from thermal freeze-out until residual annihilations no longer influence the final relic abundance. Chemical decoupling in the sense of an $\mathcal{O}(1)$ departure from chemical equilibrium occurs at approximately $T/\mDM \approx 1/25$. This places the dark matter particles clearly in a non-relativistic regime. In contrast, kinetic equilibrium persists longer, such that the typical momenta of the heavy particles remain of order $\sqrt{\mDM T}$.
In this regime, dark matter pairs are near threshold and move with non-relativistic relative velocities, $v_{\hbox{\scriptsize rel}} \sim \sqrt{T/\mDM} \ll 1$. This implies a hierarchy of relevant energy scales: the {\it hard} scale $\mDM$, and two smaller scales $\mDM v_{\hbox{\scriptsize rel}}$ and $\mDM v_{\hbox{\scriptsize rel}}^2$. Close to threshold, a fermion–antifermion pair can form either a Coulombic scattering state or a bound state.
For Coulombic bound states in the center-of-mass frame, the typical relative momentum is set by the inverse Bohr radius, $1/a_0\sim \mDM\alpha$, which we refer to as the {\it soft} scale. The corresponding binding energy, as well as the kinetic energy of a pair, is of order $\mDM\alpha^2$, defining the {\it ultrasoft} scale; more generally, we classify as ultrasoft all scales below $\mDM\alpha$.

As mentioned in the introduction, we adopt the framework of non-relativistic effective field theories (NREFTs), which enables a systematic computation of non-relativistic observables at both zero and finite temperature.
In the present paper, we do not review the NREFT framework as applied to dark matter freeze-out in detail and instead refer the reader to earlier studies \cite{Biondini:2021ccr,Biondini:2021ycj,Biondini:2023zcz,Biondini:2024aan,Biondini:2025jvp}. Rather, our focus is on collecting and discussing the relevant interaction rates that enter the Boltzmann equation for dark matter and ultimately determine the relic density under different approximations. This is the subject of sections~\ref{sec:rates_ann_dec} and \ref{sec:rates_bsf_bsd}.


The discussion that follows is also relevant for scenarios in which the actual dark matter particle is a gauge singlet, both under the SM and the dark sector gauge group, while an accompanying state carries gauge interactions in an extended dark sector. Such configurations naturally arise in coannihilation scenarios, for instance in supersymmetric models \cite{Ellis:2014ipa,Mitridate:2017izz}, scalar extensions of the SM \cite{Biondini:2017ufr,Mitridate:2017izz}, and in simplified $t$-channel dark matter frameworks \cite{Arina:2025zpi,Garny:2021qsr,Biondini:2018ovz,Bollig:2021psb,Biondini:2019int}.
\subsection{Annihilation cross section and bound-state decay width}
\label{sec:rates_ann_dec}
At energies well below the heavy mass scale $\mDM$, the relevant degrees of freedom are non-relativistic dark fermions and antifermions, low-energy dark photons, and light fermions $f_i$. The appropriate NREFTs are obtained by integrating out modes of order $\mDM$. The resulting theory is analogous to NRQED and can be organized as an expansion in $1/\mDM$ and $\alpha$.

At leading non-trivial order, the effective Lagrangian contains bilinear terms for the heavy fields and local four-fermion operators encoding annihilation processes,
\begin{eqnarray}
\mathcal{L}_{\textrm{NRQED}_{\textrm{DM}}}\!\!&=&\!\mathcal{L}^{\text{bilinear}}_{\textrm{NRQED}_{\textrm{DM}}}
        - \frac{1}{4} F^{\mu \nu}F_{\mu \nu}
       +\frac{d_2}{M^2} F^{\mu \nu} \bm{{\rm{D}}}^2 F_{\mu \nu} \nonumber
       \\
       \!&\hspace{0.5cm}+&\!\frac{d_s}{\mDM^2} \psi^\dagger \chi \, \chi^\dagger \psi
       +\frac{d_v}{\mDM^2} \psi^\dagger \, \bm{\sigma} \, \chi \cdot \chi^\dagger \, \bm{\sigma} \, \psi
       + \sum_{i=1}^{n_f}\bar{f}_i i \slashed{D} f_i \, .
\label{NREFT_lag}
\end{eqnarray}
Here, $\mathcal{L}^{\text{bilinear}}_{\textrm{NRQED}_{\textrm{DM}}}$ stands for the usual particle and antiparticle non-relativistic Lagrangian, that can be read off, e.g. refs~\cite{Bodwin:1994jh,Biondini:2023zcz}. At the lowest order in the coupling,
the heavy $X \bar{X}$ pair in an S-wave annihilates either into two dark photons or into a light
fermion-antifermion pair, respectively for a spin-singlet and spin-triplet configuration.
The imaginary parts of the Wilson coefficients $d_s$ and $d_v$ encode the short-distance annihilation of an $X\bar{X}$ pair into dark photons and light fermions respectively. At leading order, they read
\begin{equation}
\text{Im}[d_s] = \pi \alpha^2 \, , \qquad
\text{Im}[d_v] = \frac{n_f}{3}\pi \alpha^2  .
\label{matching_coeff}
\end{equation}
We stress that, at the order $1/\mDM^2$, only particle-antiparticle annihilation in the S-wave are accounted for  (particle-antiparticle P-wave annihilation, as well as velocity or $\alpha$ corrections to the S-wave, arise at order $1/\mDM^4$ \cite{Bodwin:1994jh}).

To describe near-threshold dynamics, at the typical ultrasoft scale $\mDM \alpha^2$, one further integrates out the soft scale $\mDM\alpha$, leading to pNRQED \cite{Pineda:1997bj}. In this formulation, the relevant degrees of freedom are fermion–antifermion pairs described by a bilocal field $\phi(t,\bm{r},\bm{R})$, interacting through a potential and ultrasoft radiation. The dynamics is governed by
\begin{eqnarray}
  \mathcal{L}_{\textrm{pNRQED}_{\textrm{DM}}}&=&   \int d^3r \; \phi^\dagger(t,\bm{r},\bm{R})
             \, \left[ i \partial_0 -H(\bm{r},\bm{p},\bm{P},\bm{S}_1,\bm{S}_2)  + g \, \bm{r} \cdot \bm{E}(t,\bm{R})\right] \phi (t,\bm{r},\bm{R})   \nonumber \\
&&-\frac{1}{4} F_{\mu \nu} F^{\mu \nu}   + \sum_{i=1}^{n_f}\bar{f}_i i \slashed{D} f_i  \, ,
\label{pNREFT_1}
\end{eqnarray}
where the Hamiltonian $H$ contains the kinetic terms, the Coulomb potential $V^{(0)}=-\alpha/r$, and local potentials encoding hard processes. The dipole interaction $\bm{r}\cdot \bm{E}$ governs ultrasoft transitions such as bound-state formation, dissociation or the transition between different bound states.

Annihilation processes are encoded in pNRQED through the imaginary part of a local potential inherited from NRQED,
\begin{equation}
\delta V^{\textrm{ann}}(\bm{r})
= - \frac{i}{\mDM^2} \delta^3(\bm{r}) \left[ 2\text{Im}(d_s) - \bm{S}^2 \big(\text{Im}(d_s)-\text{Im}(d_v)\big) \right] .
\end{equation}
Projecting this operator onto different two-particle states allows one to systematically extract both annihilation cross sections for scattering states and decay widths for bound states.
For scattering states, one obtains
\begin{eqnarray}
(\sigma_{\hbox{\scriptsize ann}} v_{\hbox{\scriptsize rel}})(\bm{p}) 
&=&  \frac{{\rm{Im}}(d_s)+3{\rm{Im}}(d_v)}{\mDM^2} \left|\Psi_{\bm{p} 0}(\bm{0})\right|^2
    =  (\sigma^{\hbox{\tiny NR}}_{\hbox{\scriptsize ann}} v_{\hbox{\scriptsize rel}})  \, S_{\hbox{\scriptsize ann}}(\alpha/v_\text{rel})\,,
\label{ann_fact_scat}
\end{eqnarray}
where $S_{\textrm{ann}}=|\Psi_{\bm{p}}(0)|^2$ is the Sommerfeld factor encoding long-distance effects, while $\sigma^{\textrm{NR}}_{\textrm{ann}}$ depends only on short-distance coefficients. At leading order, using the expressions in~\eqref{matching_coeff}, one finds, for the annihilation cross section and Sommerfeld factor, the following expressions 
\begin{eqnarray}
        \sigma^{\hbox{\tiny NR}}_{\hbox{\scriptsize ann}} v_{\hbox{\scriptsize rel}}=(1+n_f)\frac{\pi\alpha^2}{\mDM^2} \, , \qquad\qquad  S_{\hbox{\scriptsize ann}}(\zeta)=\frac{2 \pi \zeta}{1-e^{-2 \pi \zeta}} \, . 
        \label{non_rel_cross_and Somm}
\end{eqnarray}
with $\zeta \equiv \frac{\alpha}{v_{\hbox{\scriptsize rel}}} $.

For two-particle states below threshold, namely bound states, the decay widths are
\begin{eqnarray}
\Gamma^{n,\hbox{\scriptsize para}}_{\textrm{ann}} &=& \frac{4 {\rm{Im}}(d_s)}{\mDM^2} \frac{|R_{n0}(0)|^2}{4\pi} \,,
\label{ann_para}\\
\Gamma^{n,\hbox{\scriptsize ortho}}_{\textrm{ann}}&=& \frac{4 {\rm{Im}}(d_v)}{\mDM^2} \frac{|R_{n0}(0)|^2}{4\pi} \,,
\label{ann_ortho}
\end{eqnarray}
where $R_{n0}(r)$ is the radial part of the S-wave ($\ell=0$) wave function with principal quantum number $n$. As we are dealing with a particle-antiparticle bound-state one has to further distinguish between spin-singlet paradarkonium and spin-triplet orthodarkonium.
At leading order in the coupling, the decay widths for the para- and orthodarkonium ground state 1S are \footnote{A contribution to  the orthodarkonium decay width appears even without light fermions when going to NLO in the coupling. Besides the additional power of $\alpha$ this is also suppressed by a large phase space factor \cite{Itzykson:1980rh} and we neglect it in our work. 
} 
\begin{equation}
    \Gamma^{\text{1S},\hbox{\scriptsize para}}_{\textrm{ann}}=\frac{\mDM \alpha^5}{2} \, , \quad \Gamma^{\text{1S},\hbox{\scriptsize ortho}}_{\textrm{ann}}=\frac{n_f}{3} \frac{\mDM \alpha^5}{2} \, .
\label{eq:Gammannihilation}    
\end{equation} 
At this order in $1/\mDM$, the corresponding decay widths of higher states $nS$ can easily be found by noticing $|R_{n0}(0)|^2= 4/(na_0)^3$.  
As a side note, we mention that the natural renormalization scale for the $\alpha$ appearing in $d_s$ and $d_v$ is of the order of
$2 \mDM$, whereas the one for the coupling in the wave function is of the order of the soft scale $\mDM \alpha$.
\subsection{Bound-state formation and bound-state dissociation}
\label{sec:rates_bsf_bsd}
The next two ingredients relevant for the description of the DM pairs in the early universe are the bound-state formation cross section and the associated dissociation width. Both observables can be computed within the framework of pNRQED, as both are processes that occur at the ultrasoft scale. More specifically, the relevant energy scale is the energy difference between a scattering and a bound state. In a thermal environment a non-trivial interplay with the temperature is also realized and this is another place where the EFT approach comes in handy, whereby one may formulate a thermal version of pNRQED \cite{Escobedo:2008sy,Escobedo:2010tu,Biondini:2025jvp}.

\begin{figure}[t!]
    \centering
    \includegraphics[scale=0.75]{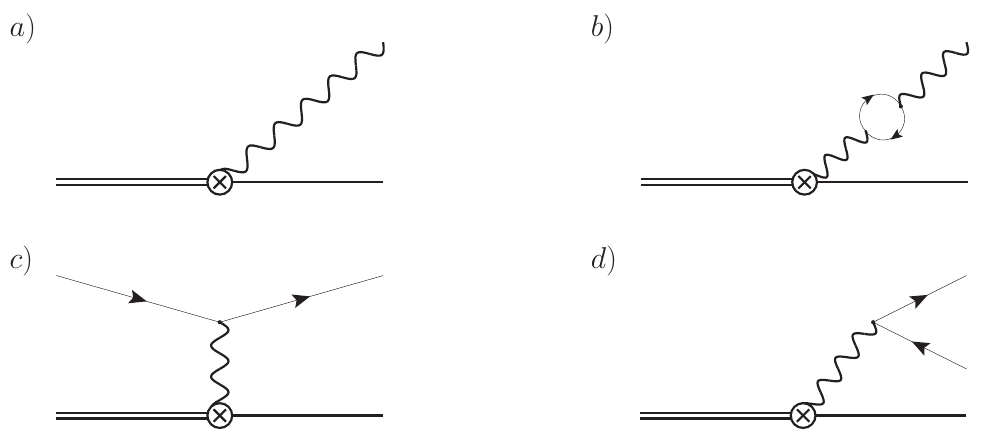}
    \caption{Bound-state formation processes.
      Solid double lines stand for dark matter pairs in a scattering state, 
      solid lines for bound states, wiggly lines for dark photons, arrowed thin lines for light plasma constituents (light fermions in our case), and  
      circled-crossed vertices for the electric dipole interactions in pNRQED$_\text{DM}$.}
    \label{fig:bsf_processes}
\end{figure}

Several processes have been identified for bound-state formation with light degrees of freedom in the dark sector in addition to the dark photon \cite{Biondini:2018pwp,Binder:2019erp,Binder:2020efn}. We show them in figure~\ref{fig:bsf_processes} and briefly describe them here. One finds: 
a) bound-state formation via radiative emission of the gauge boson;
process b), which is the same radiative emission process as a), 
however with a quantum correction for the gauge boson; c) bound-state formation via inelastic collisions of a scattering state with a light fermion of the thermal bath that turns it into a bound state; d) the decay of the emitted gauge boson into a pair of light fermions. The
reverse processes correspond to bound-state dissociation, where a) and b) can be referred
to as photo-dissociation and c) traces back to the Landau damping in a QED plasma.
In addition, bound-states can transition to one another by absorbing or emitting a boson and by scattering on light bath particles. 

The expressions for these three classes of processes (bound-state formation, bound-state dissociation, and bound-to-bound transitions) are collected below and serve as a reference for the remainder of this work.

The leading-order (LO) bound-state formation cross section due to the radiative capture process reads~\cite{vonHarling:2014kha,Biondini:2023zcz}
\begin{equation}
\begin{aligned}
(\sigma_{\text{bsf}} \, v_{\hbox{\scriptsize rel}})^{\text{LO}}(\bm{p}) 
= \sum_n (\sigma^n_{\text{bsf}} \, v_{\hbox{\scriptsize rel}})^{\text{LO}}(\bm{p})
= \frac{4}{3}\alpha(\mu)\sum_n |\langle n|\bm{r}|\bm{p}\rangle|^2 (\Delta E^p_n)^3 [1+n_{\text{B}}(\Delta E^p_n)] \, ,
\end{aligned}
\label{bsf_final_LO}
\end{equation}
where we have defined the energy difference between a scattering and bound states as
\begin{equation}
\Delta E_{n}^{p} 
\equiv E_p-E_n = 
\frac{\mDM}{4}v_{\textrm{rel}}^2\left(1+\frac{\alpha^2}{n^2v_{\textrm{rel}}^2}\right) \, .
\label{energy_photon_bsf}
\end{equation}
The transition matrix element from a scattering to a bound state due to a dipole transition is denoted $\langle n|\bm{r}|\bm{p}\rangle$ and $n_B(\Delta E_{n}^{p})$ is the dark photon occupation number at the transition energy which accounts for the Bose enhancement of process in the presence of a radiation background.
At next-to-leading order (NLO) in the coupling the bound-state formation cross section can be written as~\cite{Binder:2020efn,Biondini:2025jvp}
\begin{equation}
\begin{aligned}
(\sigma_{\text{bsf}} \, v_{\hbox{\scriptsize rel}})^{\text{LO}+\text{NLO}}(\bm{p}) 
&= \sum_n(\sigma^n_{\text{bsf}} \, v_{\hbox{\scriptsize rel}})^{\text{LO}}(\bm{p})\bigg\{1+\frac{n_f}{\pi}\alpha
\left[\mathcal{X}_1(\Delta E^p_n,\mu)
+\mathcal{X}_2(\Delta E^p_n/T)\right]\bigg\} \, ,
\end{aligned}
\label{bsf_final_fixed_NLO}
\end{equation}
where the dimensionless functions $\mathcal{X}_1$ and $\mathcal{X}_2$ are given by~\cite{Binder:2020efn,Biondini:2025jvp}
\begin{align}
&\mathcal{X}_1(\Delta E^p_n,\mu) = \ln{\left(\frac{4(\Delta E^p_n)^2}{\mu^2}\right)}-\frac{10}{3} \, ,
\label{dimless_x1}
\\
&\mathcal{X}_2(u)=\frac{2}{u^3}\int_0^{\infty}  \frac{dt}{e^{t}+1}\left[(2t^2+u^2)\ln{\left|\frac{t+u}{t-u}\right|}+2tu\ln{\left|\frac{t^2-u^2}{u^2}\right|}-2ut\right] ,
\label{dimless_x2}
\end{align}
with the dimensionless integration variable defined as $t\equiv |\bm{k}_f|/T$, where $\bm{k}_f$  is the three momentum of the light fermion. 
We note that the scale dependence in $\mathcal{X}_1$  cancels against the $\mu$ dependence of the coupling appearing in the LO expression \cite{Binder:2020efn,Biondini:2023zcz}. This holds for all the following observables. 

The dissociation width can be formulated analogously to the bound-state formation cross section. In principle, it is not necessary to compute it directly since the averaged dissociation rate, which is the quantity that enters the Boltzmann equation in the following section, can also be derived from the averaged bound-state formation cross section using the principle of detailed balance. Nevertheless, we give explicit expressions here for completeness. 
At leading order in $\alpha$, the photo-dissociation width reads
\begin{equation}
\Gamma^{n, \scalebox{0.6}{\text{LO}}}_{\textrm{bsd}} 
=2\int\frac{d^3p}{(2\pi)^3}\, \sigma^{n,\scalebox{0.6}{\text{LO}}}_{\hbox{\scriptsize ion}}(\bm{p}) n_{\textrm{B}}(|\bm{p}|)
= \frac{4}{3}\alpha \int \frac{d^3p}{(2\pi)^3} \, (\Delta E^p_n)^3 \, n_{\textrm{B}}(\Delta E^p_n) \, |\langle n|\bm{r}|\bm{p}\rangle|^2 \, ,
\label{bsd_width_LO}
\end{equation}
which implicitly defines the bound-state dissociation cross section $\sigma^{n,\scalebox{0.6}{\text{LO}}}_{\hbox{\scriptsize ion}}(\bm{p})$. Note, that the dependence on the final state momentum $p$ encodes the dependence on the energy of the incoming photon which is related to it through eq.~\eqref{energy_photon_bsf}. An equivalent formulation expresses the dissociation rate with a bound-state dissociation cross section, see e.g.~\cite{Biondini:2023ksj}.
Going to NLO the rate reads
\begin{equation}
\begin{aligned}
\Gamma_\text{bsd}^{n, \scalebox{0.6}{\text{LO}+\text{NLO}}}
&= 2\int\frac{d^3p}{(2\pi)^3}\sigma^{n,\scalebox{0.6}{\text{LO}}}_{\hbox{\scriptsize ion}}(\bm{p}) n_{\textrm{B}}(|\bm{p}|)\bigg\{1+\frac{n_f}{\pi}\alpha \left[\mathcal{X}_1(\Delta E^p_n,\mu_{\text{us}})+\mathcal{X}_2(\Delta E^p_n/T)\right]\bigg\} \, ,
\end{aligned}
\label{bsd_final_fixed_NLO}
\end{equation}
where the functions $\mathcal{X}_1$ and $\mathcal{X}_2$ are the same as above.

The last class of processes involves transitions between bound states. These reactions become relevant when one includes excited states. The inclusion of bound-to-bound transitions can become computationally demanding when many states are considered, and has been the subject of recent developments of semi-automated approaches \cite{Becker:2026icc,Binder:2025daq}. The excitation and de-excitation widths at LO read
\begin{align}
\Gamma^{n, \scalebox{0.6}{\text{LO}}}_{\textrm{de-ex.}} &= \frac{4}{3}\alpha \sum_{n', E_{n'}<E_n}\left(\Delta E_{n'}^{n}\right)^3 \, \left(1+n_{\text{B}}\left(\Delta E_{n'}^{n}\right)\right) \, \left| \langle n'|  \bm{r}  | n \rangle \right|^2 \, ,
\label{gamma_de-excitation_LO}
\\
\Gamma^{n, \scalebox{0.6}{\text{LO}}}_{\textrm{ex.}} &= \frac{4}{3}\alpha \sum_{n', E_{n'}>E_n} (\Delta E_{n}^{n'})^3 \, n_{\text{B}}(\Delta E_{n}^{n'}) \, \left| \langle n' | \bm{r} | n \rangle \right|^2 \, .
\label{gamma_excitation_LO}
\end{align}
and they appear in the NLO expressions, which can be easily constructed by taking care of the sum over the bound states in combination with the energy difference depending functions $\mathcal{X}_1$ and $\mathcal{X}_2$.  
In the case of bound-to-bound transitions, the energy differences between levels are simply
\begin{align}
\Delta E^n_{n'} \equiv E_n - E_{n'} = \frac{\mDM \alpha^2}{4} \left(\frac{1}{{n'}^2}-\frac{1}{{n}^2}\right) \, , \qquad 
\Delta E^n_{n'} = - \Delta E^{n'}_{n} \, ,
\end{align}
since there is no scattering state that could provide or absorb additional kinetic energy.

\section{Reduced decay width and approximate particle interaction rates}
\label{sec:effective_cross_approximate_rates}
In this section we discuss the Boltzmann equation used to describe the dark matter evolution when near-threshold effects are included. 
Our goal is to briefly review the framework commonly adopted to construct the Boltzmann equation, and to clarify how the decay and dissociation widths, as well as the bound-state formation processes discussed in section~\ref{sec:rates}, enter the rate equation. Once this is done we turn to outline simple but accurate approximations to the various quantities that enter the Boltzmann equation. 

Including bound states in the network of Boltzmann equations generally leads to a coupled system that can be cumbersome to solve. In this work, we adopt an approximation, which was first introduced in ref.~\cite{Ellis:2015vaa} and is widely used in the literature, based on an effective treatment of dark matter bound states. 
If the Hubble rate $H \ll \text{Max}[\Gamma^{n,\hbox{\scriptsize para\,(ortho)}}_{\textrm{ann}}, \Gamma^{n}_{\textrm{bsd}}]$,  the time derivative of the bound state number density can be neglected which turns one of the Boltzmann equations into an algebraic relation. This is the case in the cosmological scenario considered here. Under this assumption, the coupled Boltzmann equations can be recast as a single evolution equation for the total dark matter density $n = n_X + n_{\bar{X}} = 2 n_X$~\cite{Ellis:2015vaa}. 

Neglecting transitions between different bound states, the resulting Boltzmann equation depends only on the density of scattering states $n$ and is governed by an effective cross section (see e.g.~\cite{Biondini:2023zcz} for a detailed derivation). This effective description incorporates annihilation from both unbound pairs and bound states, as well as bound-state formation and dissociation processes. 
Recent works have further explored the validity and possible extensions of this approach when transitions among bound states are included in the Boltzmann system \cite{Garny:2021qsr,Binder:2021vfo}. The structure of the effective cross section can be kept, while a matrix involving combinations of decay, dissociation, and transitions widths of bound state also enters. 

Under these assumptions, the Boltzmann equation takes the form
\begin{equation}
    (\partial_t + 3H) n = - \frac{1}{2}\langle \sigma_{\textrm{eff}} \, v_{\textrm{rel}} \rangle (n^2-n^2_{\textrm{eq}}) \,,
    \label{Boltzmann_eq_eff}
\end{equation}
where $n_\textrm{eq} = 2n_{X,\textrm{eq}}=4(\mDM T/2 \pi)^{3/2}e^{-\mDM/T}$, $H$ is the Hubble rate of the Universe,\footnote{In radiation domination the Hubble rate takes the form $H = \sqrt{4 \pi^3 g_{\rmi{eff}}(T)/45} T^2 / m_{\rmii{Pl}}$, where $g_{\rmi{eff}}$ are the temperature-dependent relativistic degrees of freedom.} and the thermally averaged {\it effective cross section} is defined as (in the no-transitions limit)
\begin{equation}
   \langle  \sigma_{\textrm{eff}} \, v_{\textrm{rel}} \rangle  =
   \langle \sigma_{\textrm{ann}} v_{\textrm{rel}} \rangle
   + \sum_{n} \left(\frac{1}{4}\langle   \sigma^n_{\textrm{bsf}} \, v_{\textrm{rel}} \rangle \, \frac{\Gamma^{n,\hbox{\scriptsize para}}_{\textrm{ann}}}{\Gamma^{n,\hbox{\scriptsize para}}_{\textrm{ann}}+\Gamma_{\textrm{bsd}}^n} +\frac{3}{4}\langle   \sigma^n_{\textrm{bsf}} \, v_{\textrm{rel}} \rangle \, \frac{\Gamma^{n,\hbox{\scriptsize ortho}}_{\textrm{ann}}}{\Gamma^{n,\hbox{\scriptsize ortho}}_{\textrm{ann}}+\Gamma_{\textrm{bsd}}^n} \right) .
    \label{Cross_section_eff}
\end{equation}
In eq.~\eqref{Cross_section_eff}, spin-triplet and spin-singlet bound states are treated on the same footing, as orthodarkonium and paradarkonium decay at the same order in the coupling. In general, the bound state dissociation has to be understood as the sum of the processes that we have described in the previous sections \ref{sec:rates_ann_dec} and \ref{sec:rates_bsf_bsd}. At LO, it only comprises photo-dissociation (with no correction to the photon propagator), whereas at NLO the bath scattering induced dissociation has to be included. 
The same applies to the bound-state formation.

In the remainder of this section we will study the behavior of these contributions and identify effective parametrizations that capture the leading dynamics. We argue that the key ingredients for a reasonably accurate determination of the relic density are the high and the low temperature limits of the rates. These can be evaluated analytically and, therefore, it is possible to construct simple, physics motivated approximating functions that reproduce all qualitative features of the full results. As we will demonstrate in this and the following section, the quantitative agreement is also very good when the coupling remains in the perturbative regime.   


\subsection{Sommerfeld enhanced annihilation}
\label{sec:approx_Somm}
As is well known, the asymptotic behavior of the thermally averaged Sommerfeld factor for an attractive potential is given by
\begin{equation}
\langle S_{\rmi{ann}} \rangle = \begin{cases}
      1 \,,  \quad  \sqrt{z}{} \ll 1/\alpha  \,,\\
       2 \sqrt{ \pi} \alpha \sqrt{z}  \, , \quad \sqrt{z} \gg 1/\alpha \, .
    \end{cases} 
    \label{eq:Sommerfeld_asymptotics}
\end{equation}
where $z =m_{\textrm{DM}}/T$.
A convenient approximation of $\langle S_{\rmi{ann}}\rangle$ that smoothly connects both limits and deviates from the full result by less than $4\%$ everywhere is
\begin{align}
    \langle S_{\text{ann}} \rangle_{\text{approx}} = 2 \sqrt{\pi } \alpha  \sqrt{z}+\left(\frac{4 \sqrt{\pi }}{3 \left(3 \alpha  \sqrt{z}+2\right)} -\frac{2}{3}
   \sqrt{\pi }\right)+ 1 \,.
   \label{approx_Sommerfeld}
\end{align}
Naively joining the limiting behaviors, i.e. omitting  the second term, incurs an error of up to $20\%$ instead. 
\begin{figure}[t!]
    \centering
    \includegraphics[width=0.45\linewidth]{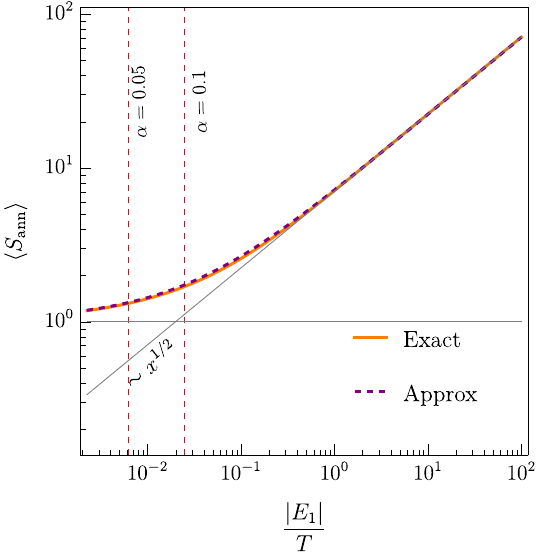}
    \includegraphics[width=0.46\linewidth]{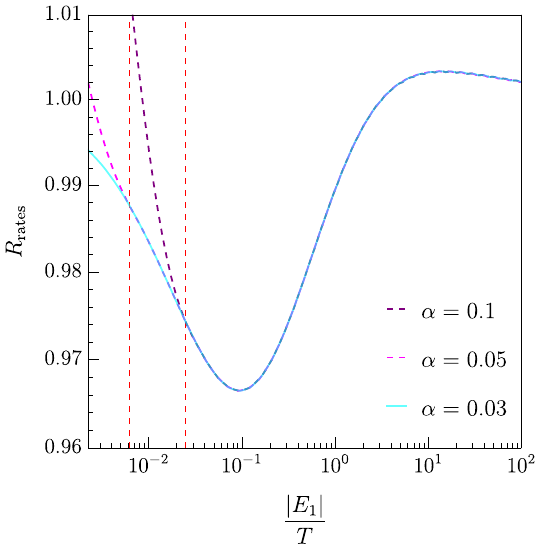}
    \caption{(Left) The numerically extracted thermally averaged Sommerfeld factor from the exact expression \eqref{non_rel_cross_and Somm}, orange-solid line, and the approximated thermally averaged Sommerfeld \eqref{approx_Sommerfeld}, purple-dashed. The asymptotic behaviors at small and large temperatures are shown. The vertical dashed lines indicate the value of $|E_1|/T$ that corresponds to a temperature of $T=m_{\textrm{DM}}/10$ for different values of the coupling $\alpha$. (Right): Ratio of the thermal average of the approximate Sommerfeld factor eq.~\eqref{approx_Sommerfeld} and the exact Sommerfeld factor \eqref{non_rel_cross_and Somm}. The vertical dashed lines indicate $T=m_{\textrm{DM}}/10$ for different values of the coupling $\alpha$ following the same style as in the left panel. }
    \label{fig:comparison_Sommerfeld}
\end{figure}

In figure~\ref{fig:comparison_Sommerfeld}, we show the Sommerfeld-enhanced annihilation cross section normalized to the non-relativistic result for $n_f=0$, namely $\sigma^{\rmii{NR}} \vrel=\pi \alpha/\mDM^2$. In analogy with the discussion of bound-state effects, we use the dimensionless variable defined by the ratio of the ground-state binding energy to the temperature, $x \equiv |E_1|/T$. 
The connection with the frequently used time variable $z=\mDM/T$ is given by
\begin{equation}
x=\frac{\alpha^2 z}{4} \, .
\end{equation}

When the approximate Sommerfeld factor is expressed in terms of $x=|E_1|/T$, rather than $z=\mDM/T$, one finds that $ \langle S_{\text{ann}} \rangle_{\text{approx}}$ does not depend on $\alpha$ (because $\alpha \sqrt{z} = 2\sqrt{x}$) in contrast to the full result. However, this is not an issue as the departure only occurs for temperatures where the system is in chemical equilibrium such that the exact value of the rate does not matter for the dynamics. 

In figure~\ref{fig:comparison_Sommerfeld}, the minimum value shown in the $x$ range corresponds to $\alpha=0.03$. More specifically, choosing the typical lower bound $z_{\rmi{min}}=10$ translates into $x_{\rmi{min}}=\alpha^2 z_{\rmi{min}}/4$. We therefore take the lower end of the displayed range to be $x = 2.25 \times 10^{-3}$, corresponding to $z_{\rmi{min}}=10$ for $\alpha=0.03$. 
For comparison, the values of $x$ corresponding to $\alpha=0.05$ and $\alpha=0.1$ are indicated by the vertical red dashed lines in the left and right panels, respectively.

\subsection{Bound-state dissociation}
\label{sec:approx_bound_state_diss}

As a first step towards a simple parametrization of the bound state effects, we recast the bound-state contribution in eq.~\eqref{Cross_section_eff} in terms of a reduced decay width. Focusing on the paradarkonium contribution as a representative example, one can rewrite it as
\begin{eqnarray}
     \langle   \sigma^n_{\textrm{bsf}} \, v_{\textrm{rel}} \rangle \, \frac{\Gamma^{n,\hbox{\scriptsize para}}_{\textrm{dec}}}{\Gamma^{n,\hbox{\scriptsize para}}_{\textrm{dec}}+\Gamma_{\textrm{bsd}}^n} = \frac{\Gamma^{n,\hbox{\scriptsize para}}_{\textrm{dec}} \, \Gamma_{\textrm{bsd}}^n}{\Gamma^{n,\hbox{\scriptsize para}}_{\textrm{dec}}+\Gamma_{\textrm{bsd}}^n} e^{\frac{|E_n|}{T}}\left( \frac{ 2\pi}{ \muDM T}\right)^{3/2} \, \equiv \Gamma^{n,\textrm{para}}_{\textrm{red}} e^{\frac{|E_n|}{T}}\left( \frac{ 2\pi}{\muDM T}\right)^{3/2} \, , \nonumber
     \\
     \phantom{s}
     \label{interm_det_bal}
\end{eqnarray} 
 where the bound-state formation cross section has been eliminated by means of the principle of detailed balance and we also introduce the reduced mass of the dark matter particle–antiparticle pair, $\mu_{\rmii{DM}}=m_{\rmii{DM}}/2$. This allows to parametrize the contribution of bound-states to the effective cross section in terms of a ``reduced'' width and a statistical factor which depends on the particle physics only through the binding energy. In analogy with the reduced mass, this quantity approaches the smaller of the two rates when a strong hierarchy exists between them. 
As the dissociation rates increase with temperature, while the decay width remains constant, we can already see that the effective contribution will be proportional to the decay width at large $T$ and the dissociation rate at low $T$. 
Equation~\eqref{interm_det_bal} makes it clear that the bound-state contribution to the effective cross section can be conveniently expressed in terms of the dissociation width. Therefore, in the following, we focus on the dissociation rates, which provide direct access to the impact of bound states. 
To streamline the notation, we now define an abbreviation for the bound-state contribution to $ \langle  \sigma_{\textrm{eff}} \, v_{\textrm{rel}} \rangle $ in eq.~\eqref{Cross_section_eff}
\begin{eqnarray}
    \Gamma^{n}_{\textrm{red}} e^{\frac{|E_n|}{T}}\left( \frac{ 2\pi}{\muDM T}\right)^{3/2}  = \mathcal{B}^{n} \, ,
    \label{def_effective_BS_rate}
\end{eqnarray}
which holds for both para- and orthodarkonium states. 

\subsubsection{LO bound-state dissociation}
\label{sec:bsd_LO}
We begin by considering the leading-order photo-dissociation process, which provides the dominant contribution to bound-state effects for values of $\alpha$ where perturbation theory is reliable. For simplicity, we focus on the ground state, although the same reasoning applies to higher bound states. Starting from eq.~\eqref{bsd_width_LO} the dissociation width for the ground state can be found by using the explicit expressions for the wave-functions.
Nevertheless, it is instructive to take an intermediate step first and take a look at the leading-order ionization cross section for the 1S state which is given by 
\begin{align}
\sigma^{1 \textrm{S}}_{\hbox{\scriptsize ion}}(|\bm{k}_\gamma|)= \alpha \frac{2^{9} \pi^2}{3} \frac{|E_1^b|^3}{M |\bm{k}|^4} \frac{e^{-\frac{4}{w_1(|\bm{k}_\gamma|)}\arctan(w_1(|\bm{k}_\gamma|))}}{1-e^{-\frac{2\pi}{w_1(|\bm{k}_\gamma|)}}}
\label{ion_1S_exact}
\end{align}
with $w_1(|\bm{k}_\gamma|) \equiv \sqrt{|\bm{k}_\gamma|/|E_1|-1}$; see e.g.~ref.~\cite{Biondini:2023zcz,Bethe:1957ncq}. This cross section is strongly peaked towards the threshold and  approximately decreases as a power law $|\bm{k}_\gamma|^{-\eta}$ with an energy dependent exponent $\eta$ that varies between $8/3$ on threshold and $7/2$ for $|\bm{k}_\gamma| \gg |E_1|$. As a consequence, the thermal averages  are dominated by photon energies close to the threshold even in the high temperature limit. In the relevant regime for the averages, i.e. for $|\bm{k}_\gamma|\lesssim \mbox{a few } |E_1|$, one can approximate 
\begin{align}\frac{e^{-4 \frac{\arctan(w_1(|\bm{k}_\gamma|))}{w_1(|\bm{k}_\gamma|)}}}{1-e^{-2 \pi/w_1(|\bm{k}_\gamma|)} } \approx \frac{1}{e^4}\left(\frac{4}{3}\frac{|\bm{k}_\gamma|}{|E_1|}-\frac{1}{3}\right) \, ,
\label{eq:Bethe_approximation}
\end{align}
as suggested in ref.~\cite{Bethe:1957ncq}. This expression is exact at threshold, reproduces the full result with an accuracy better than $1.5\%$ for $|\bm{k}_\gamma|\le 5|E_1|$, and remains within a factor of two even up to $|\bm{k}_\gamma|=100 \times |E_1|$. Since the thermal average is dominated by photon energies close to threshold, this approximation provides an excellent description of the dissociation width while considerably simplifying the subsequent analytic treatment.

We now bring the dissociation width in a simple form that depends explicitly on the ratio of binding energy to temperature,
\begin{eqnarray}
    \Gamma^{1 \textrm{S}}_{\textrm{bsd}} 
    = \frac{2^7}{3} \alpha^3 \, T \left(  \frac{|E_1|}{T} \right)^2 \, 
   \mathcal{F}^{1 \textrm{S}}\left( \frac{|E_1|}{T}\right)   \, ,
   \label{exact_diss_widths_1S}
\end{eqnarray}
where
\begin{eqnarray}
    \mathcal{F}^{1 \textrm{S}}(x) = \int_{x}^{\infty} d y \, \frac{1}{y^2 (e^y-1)} \frac{e^{-4 \frac{\arctan(w_1(x,y))}{w_1(x,y)}}}{1-e^{-2 \pi/w_1(x,y)}} \, ,
    \label{integral_diss_widths_1S}
\end{eqnarray}
and $w_1(x,y)=\sqrt{y/x-1}$, where $y=|\bm{k}_\gamma|/T$ and $x=|E_1|/T$.
%
Although the full expression involves a non-trivial integral, its limiting behavior can be understood analytically with the approximation in \eqref{eq:Bethe_approximation}, once the thermal weight factor is expanded as appropriate in the two limits.
For $T \ll |E_1|$, the dissociation rate is exponentially suppressed, $\Gamma_{\textrm{bsd}} \propto \alpha^3 T e^{-|E_1|/T}$, reflecting the Boltzmann suppression of energetic photons in the thermal bath. In the opposite limit, $T \gg |E_1|$, the rate grows linearly with temperature, $\Gamma_{\textrm{bsd}} \propto \alpha^3 T$.

Motivated by these limits, one can construct a simple expression that captures the dominant temperature dependence over the entire range. A convenient approximation, which reproduces the correct low-temperature behavior and provides a good description at high temperatures, is
\begin{align}
    \Gamma^{1 \textrm{S}}_{\textrm{bsd}} \approx \frac{128}{3 e^4}\,\alpha^3 T \, e^{-|E_1|/T} \, .
    \label{1_to_2_photo_approx_anyT}
\end{align}
The exponential factor encodes the Boltzmann suppression characteristic of the threshold regime, while the prefactor $\sim \alpha^3 T$ reflects the scaling of the underlying electric-dipole transition. The numerical coefficient is found by matching to the analytic result in the low temperature limit.\footnote{Exact agreement in the opposite limit $T\gg E_1$ would be found by rescaling this result for $\Gamma^{1 \textrm{S}}_{\textrm{bsd}}$ by a factor $7/6$ and dropping the exponential which tends to 1 here anyways. As the exact dependence in this limit turns out to be less important for the overall dynamics we omit this correction here.}

It is worth noting that the exponential suppression in eq.~\eqref{1_to_2_photo_approx_anyT} cancels against the thermal enhancement appearing in the reduced rate, see~eq.~\eqref{interm_det_bal}. As a result, the bound-state contribution to the effective cross section scales as $1/\sqrt{T}$ for $T \ll |E_1|$, cf.~eq.~\eqref{BR_power_LO_para}, similarly to the Sommerfeld-enhanced annihilation cross section in this regime.
As both are $\mathcal{O}(\alpha^3)$, this implies that two contributions scale exactly the same in this regime and their relative importance is reduced to a constant $\mathcal{O}(1)$ factor.

The approximate expression above is useful in practical computations, as it avoids the numerical evaluation of $\mathcal{F}(x)$ while retaining the correct parametric behavior. Using eq.~\eqref{interm_det_bal}, the effective rate describing the bound-state contribution to dark matter annihilation can be written in terms of $x=|E_1|/T$ as
\begin{eqnarray}
 \mathcal{B}^{\rmii{1S},\rmi{para}}_{\rmi{approx}} = \frac{\pi \alpha^2}{\mDM^2} \frac{2^5 \sqrt{\pi} \,  64}{3 e^4} \frac{x^{\frac{1}{2}}}{1+ \frac{64}{3e^4}\frac{e^{-x}}{x}} \, .
\label{effective_rate_para_approx}
\end{eqnarray}

\begin{figure}[t!]
    \centering
        \includegraphics[width=0.495\linewidth]{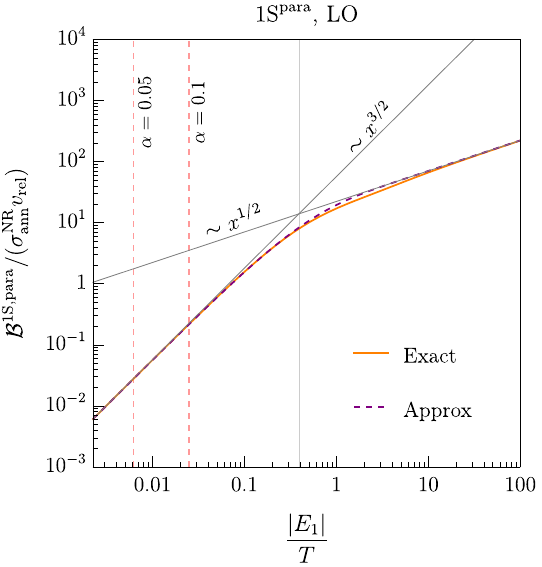}
        \includegraphics[width=0.48\linewidth]{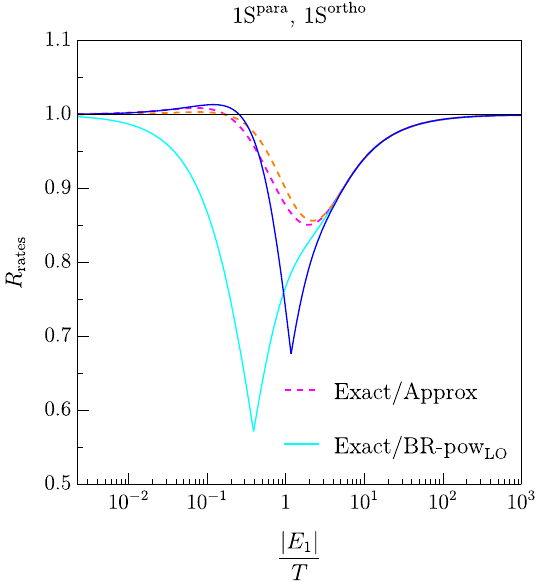}
    \caption{(Left) Effective rate capturing the bound-state contribution to dark matter annihilation, normalized to the hard cross section $\sigma^{\hbox{\tiny NR}}_{\hbox{\scriptsize ann}} v_{\hbox{\scriptsize rel}}=\pi \alpha^2/\mDM^2$ for $n_f=0$. The orange line corresponds to the full pNRQED result, while the purple dashed line is obtained using the approximation in eq.~\eqref{1_to_2_photo_approx_anyT}. The asymptotic behaviors at large and small $x$ are also indicated. (Right) We show the ratio of the bound-state contribution in eq.~\eqref{def_effective_BS_rate}, computed using the dissociation width in eq.~\eqref{exact_diss_widths_1S}, to the results obtained with the approximate width in eq.~\eqref{1_to_2_photo_approx_anyT} and the power-law expression in eq.~\eqref{BR_power_LO_para}, represented by the magenta-dashed and cyan solid lines, respectively for paradarkonium. The orange dashed and solid blue lines represent the orthodarkonium counterparts with $n_f=1$. }
    \label{fig:comparison_photo_diss}
\end{figure}

As an even more aggressive simplification one may construct a broken power-law approximation by matching the asymptotic behaviors of eq.~\eqref{effective_rate_para_approx} at small and large $x$. This leads to the piecewise form 
\begin{equation}
\mathcal{B}^{\rmii{1S},\rmi{para}}_{\rmi{BR\,pow},\rmii{LO}} = \begin{cases}
      2^5 \sqrt{\pi} \frac{\pi \alpha^2}{\mDM^2}  x^{\frac{3}{2}} \,,  \quad  x \leq  \frac{64}{3e^4} \,,\\[4 mm]
      2^5 \sqrt{\pi} \frac{\pi \alpha^2}{\mDM^2}  \frac{64}{3 e^4} \, x^{\frac{1}{2}} \, , \quad x > \frac{64}{3e^4} \, .
    \end{cases} 
    \label{BR_power_LO_para}
\end{equation}
This approximation is cruder than eq.~\eqref{effective_rate_para_approx} and systematically overestimates the bound-state contribution for leading order particle rates. A comparison between the full pNRQED result (solid-orange), the smooth approximation (purple-dotted), and the broken power-law form (solid-gray) is shown in figure~\ref{fig:comparison_photo_diss}, left panel. The minimum value for the $x$ range is taken for $\alpha=0.03$. More specifically, the commonly used variable $z=\mDM/T$, with $z_{\rmi{min}}=10$ corresponds to $x_{\rmi{min}}=\alpha^2 z_{\rmi{min}}/4$. We indicate with dashed vertical lines the corresponding values of $|E_1|/T$ for which $z=10$ for two other benchmark values of $\alpha=0.05$ and $\alpha=0.1$. We stress that, when normalizing the effective rate by $\sigma^{\hbox{\tiny NR}}_{\hbox{\scriptsize ann}} v_{\hbox{\scriptsize rel}} \equiv \pi \alpha^2/\mDM^2$, no residual $\alpha$ dependence remains. This holds true also for $\mathcal{B}^n$ as build with the exact rates. 

In order to show clearly the discrepancy induced by the approximate rates, in the right panel of figure~\ref{fig:comparison_photo_diss}, we divide the exact expression $\mathcal{B}^{\rmii{1S},\rmi{para}}$ by the approximate results: the magenta dashed line corresponds to $\mathcal{B}^{\rmii{1S},\rmi{para}}_{\rmi{approx}}$, while the cyan solid line corresponds to $\mathcal{B}^{\rmii{1S},\rmi{para}}_{\rmi{BR\,pow}}$.  
The approximate dissociation width leads to a mild underestimation of the effective rate for $x \lesssim 0.1$, followed by an overestimation at larger values of $x$. The maximum discrepancy is about $15\%$ at $x \simeq 2$, after which the result gradually approaches the power-law behavior $\propto x^{1/2}$ for $x \gtrsim 10$. 
The broken power-law approximation, on the other hand, systematically overestimates the effective rate, and does so more significantly than $\mathcal{B}^{\rmii{1S},\rmi{para}}_{\rmi{approx}}$. However, for $x \gtrsim 2$, the difference between the smooth approximation and the broken power-law becomes practically negligible. 

The same discussion applies for the bound-state contribution of ortho-darkonium states to the effective annihilation cross section, which are non-vanishing for $n_f \geq 1$. The corresponding approximate rate and the broken power law read respectively 
\begin{align}
& \mathcal{B}^{\rmii{1S},\rmi{ortho}}_{\rmi{approx}} = \frac{\pi \alpha^2}{\mDM^2}   \frac{2^5 \sqrt{\pi} 64}{3 e^4} \frac{n_f \, x^{\frac{1}{2}}}{n_f+ \frac{64}{e^4 }\frac{e^{-x}}{x}} \, ,
\label{effective_rate_ortho_approx}
\\
&\mathcal{B}^{\rmii{1S},\rmi{ortho}}_{\rmi{BR\,pow},\rmii{LO}} = \begin{cases}
      2^5 \sqrt{\pi} \frac{\pi \alpha^2}{\mDM^2}  \frac{n_f}{3} x^{\frac{3}{2}} \,,  \quad  x \leq  \frac{64}{e^4 \, n_f} \,,\\[4 mm]
      2^5 \sqrt{\pi} \frac{\pi \alpha^2}{\mDM^2}  \frac{64}{3 e^4} \, x^{\frac{1}{2}} \, , \quad x > \frac{64}{e^4 \, n_f} \, .
    \end{cases} 
    \label{BR_power_LO_ortho}
\end{align}
Here we have used the LO dissociation width as for the ortho-darkonium case. Indeed, as mentioned in section~\ref{sec:rates}, light-fermion species trigger the bound-state decay for darkonium in a spin-triplet configuration at leading order in the coupling. In the right panel of figure~\ref{fig:comparison_photo_diss}, we show the discrepancy between the $\mathcal{B}^{\rmii{1S},\rmi{ortho}}$ as obtained with the exact rate and its approximations for $n_f=1$. The curves are the orange dashed and solid blue respectively for the ratio involving the approximate rate \eqref{effective_rate_ortho_approx} and the power law \eqref{BR_power_LO_ortho}. 

\subsubsection{NLO bound-state dissociation}
\label{sec:bsd_NLO}
A similar strategy for the extraction of approximate rates can be applied when including the NLO contributions. Although the individual rates become significantly larger than their LO counterparts at high temperatures, $T \gg |E_1|$, bound-state formation and dissociation combine in the effective rate in such a way that the net effect is only a moderate, order-one enhancement, depending on the value of $\alpha$. 

As shown in eq.~\eqref{bsd_final_fixed_NLO}, the NLO correction consists of two terms. The first is a temperature-independent contribution, which cannot be further simplified. The renormalization scale appearing in the logarithm can be chosen as $\mu \sim \Delta E_n^p$, so that this term reduces to a constant\footnote{Another possible choice is $\mu = \mDM v^2/4$, as done e.g. in \cite{Biondini:2025jvp}.}
\begin{equation}
\mathcal{X}_1 = \ln(4) - \frac{10}{3} \simeq -0.65 \, .
\end{equation}

The second term, $\mathcal{X}_2$, provides a sizable enhancement of the individual rates in the high-temperature regime, $T \gg |E_1|$. We find that its behavior in the two limiting regimes is
\begin{align}
      &\mathcal{X}_{2}(y)\big|_{y \gg 1}= \frac{14 \pi^4}{45 y^4} \, , \\
      &\mathcal{X}_{2}(y)\big|_{y \ll 1} = \frac{\pi^2}{3 y^2} \left[\ln (1/y^2) +\kappa \big)\right] \, ,
      \label{limit_x2_small}
\end{align}
where 
\begin{eqnarray}
    \kappa = 1- 2 \gamma_{\textrm{E}} - 2 \ln 2 + \frac{6}{\pi^2}(\gamma'_1(1/2)-\gamma'_1(1)) \simeq 2.091 \, 
\end{eqnarray}
and
$\gamma'_1(a) \equiv d \gamma_1(a)/da$ denotes the derivative of the generalized Stieltjes constant $\gamma_n(a)$ for $n=1$. 

We wrote $\mathcal{X}_2(y)$ with $y=|\bm{k}_\gamma|/T$ being the variable in the thermal average of the dissociation width \eqref{exact_diss_widths_1S}. The integral over $y$ will affect the $\mathcal{X}_2$ term only, since $\mathcal{X}_1$ does not depend on the photon energy given our choice of the renormalization scale. In the high and low temperature limit the integral can be performed analytically with the same strategy as for the LO case. 
\begin{figure}
    \centering
    \includegraphics[width=0.48\linewidth]{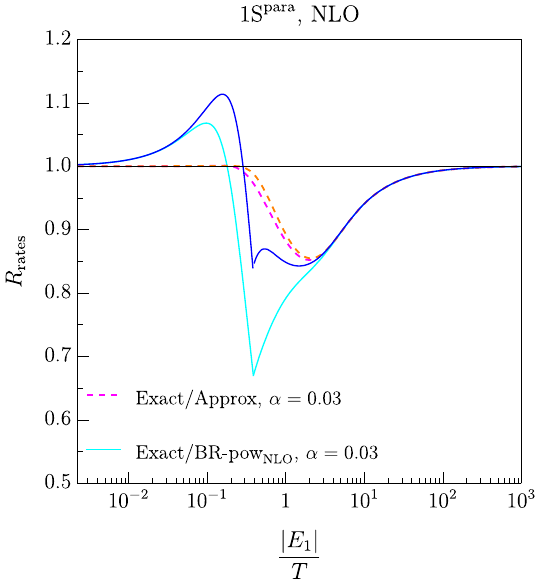}
    \includegraphics[width=0.48\linewidth]{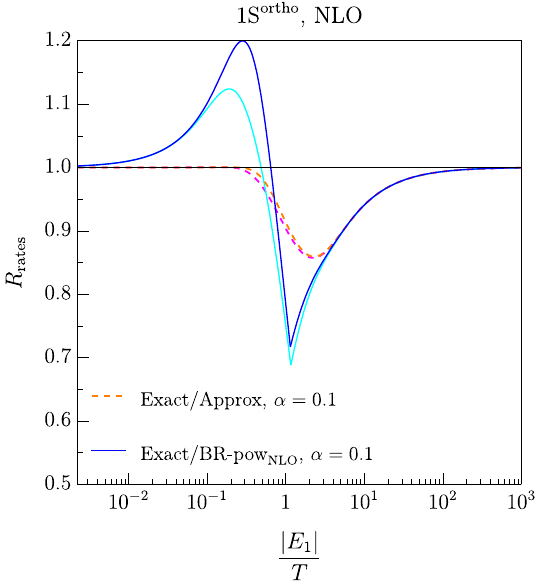}
    \caption{(Left) Ratio of the bound-state contribution $\mathcal{B}^{\rmii{1S},\rmi{para}}$, computed using the dissociation width in eq.~\eqref{bsd_final_fixed_NLO}, to the corresponding results obtained with the approximate width in eq.~\eqref{approximate_diss_width_NLO} and the broken power-law expression in eq.~\eqref{BR_power_NLO_para}. For $\alpha=0.03$, the approximate and broken power-law results are shown by the magenta dashed and cyan solid lines, respectively. The orange dashed and blue solid lines correspond instead to $\alpha=0.1$; their legend is shared with the right panel showing the orthodarkonium case. 
(Right) Same as the left panel, but for the orthodarkonium contribution $\mathcal{B}^{\rmii{1S},\rmi{ortho}}$. In this case, the number of light fermions is fixed to $n_f=1$.}
    \label{fig:comparison_photo_diss_NLO}
\end{figure}
One obtains the analogue of the approximate dissociation width in eq.~\eqref{1_to_2_photo_approx_anyT}, namely
\begin{equation}
       \Gamma^{\text{1S}}_{\textrm{bsd}} \simeq \frac{128}{3 e^4} \alpha^3 T e^{-|E_1|/T} \left( 1+ \frac{\alpha}{\pi} n_f \big(\mathcal{X}_1+\bar{\mathcal{X}}_2(x)\big)\right) \equiv  \frac{128}{3 e^4} \alpha^3 T e^{-|E_1|/T} \left( 1+ \bar{\mathcal{X}}(x)\right)\, ,
       \label{approximate_diss_width_NLO}
\end{equation}
where the barred quantities denote the averaged over the thermal photon momentum. 
By combining the asymptotic behavior in the limits $T \ll |E_1|$ and 
$T \gg |E_1|$, one can construct a simple interpolating expression that captures the dominant temperature dependence over the full range. A convenient form is
\begin{eqnarray}
 \bar{\mathcal{X}}_2(x) =  \frac{14\pi^4}{45}\, \left( x^4  + \frac{168 \pi^2}{65} \frac{x^2}{2\log(1+1/x)+\bar{\kappa}}\right)^{-1}
 \label{barX2_averaged_NLO}
\end{eqnarray}
where $\bar{\mathcal X}_2(x)$ denotes the result after performing the integral over $y$ and $\bar{\kappa}=\kappa -55/78$. We note in passing  that a constant correction has been added to the argument of the $\log$ term by hand in eq.~\eqref{barX2_averaged_NLO}. This  prevents an unphysical negative contribution for $x>1$. 

With this result at hand, the approximation for $\mathcal{B}^{\rmii{1S},\rmi{para}}_{\rmi{approx}}$ in eq.~\eqref{effective_rate_para_approx} can be systematically improved by including the correction factor $\bar{\mathcal{X}}(x)$. An analogous improvement can also be implemented for the orthodarkonium contribution in eq.~\eqref{effective_rate_ortho_approx}.  

As a final ingredient in our set of approximate rates, we discuss the broken power-law approximation including NLO corrections. When incorporating these corrections into the broken power-law expressions of eq.~\eqref{BR_power_LO_para} and eq.~\eqref{BR_power_LO_ortho}, one finds the following behavior. In the limit $x \ll 1$, the functional form remains unchanged. Physically, this follows from the fact that, at high temperatures, the dissociation width dominates over the decay width, a hierarchy that is further enhanced by the large thermal corrections in eq.~\eqref{limit_x2_small}. As a consequence, the dependence on $n_f$ and $\alpha$ drops out from the functional form of $\mathcal{B}^n$ in this regime.
In contrast, in the limit $x \gg 1$, the rate acquires an overall correction factor depending on both $n_f$ and $\alpha$. In this regime, only the constant contribution from $\mathcal{X}_1$ survives, since the $\mathcal{X}_2$ contribution decreases much faster with temperature than the leading-order terms in the dissociation width. Consequently, the transition point between the two asymptotic regimes is shifted, and the NLO-corrected broken power-law approximation reads
\begin{equation}
\mathcal{B}^{\rmii{1S},\rmi{para}}_{\rmi{BR\,pow},\rmii{NLO}} = \begin{cases}
      2^5 \sqrt{\pi} \frac{\pi \alpha^2}{\mDM^2}  x^{\frac{3}{2}} \,,  \quad  x \leq  \frac{64}{3e^4} \left( 1-0.65\, n_f\frac{\alpha}{\pi}\right) \,,\\[4 mm] 
      2^5 \sqrt{\pi} \frac{\pi \alpha^2}{\mDM^2}  \frac{64}{3 e^4} \, x^{\frac{1}{2}} \,  \left( 1-0.65\, n_f\frac{\alpha}{\pi}\right) \, ,\quad x > \frac{64}{3e^4} \left( 1-0.65\, n_f\frac{\alpha}{\pi}\right) \, ,
    \end{cases} 
\label{BR_power_NLO_para}
\end{equation}
for the paradarkonium, whereas as follows for the orthodarkonium 
\begin{equation}
\mathcal{B}^{\rmii{1S},\rmi{ortho}}_{\rmi{BR\,pow},\rmii{NLO}} = \begin{cases}
      2^5 \sqrt{\pi} \frac{\pi \alpha^2}{\mDM^2}  \frac{n_f}{3} \, x^{\frac{3}{2}} \,,  \quad  x \leq  \frac{64}{e^4 n_f} \left( 1-0.65\, n_f\frac{\alpha}{\pi}\right) \,,\\[4 mm]
      2^5 \sqrt{\pi} \frac{\pi \alpha^2}{\mDM^2}  \frac{64}{3 e^4} \, x^{\frac{1}{2}} \,  \left( 1-0.65\, n_f\frac{\alpha}{\pi}\right) \, ,\quad x > \frac{64}{e^4 n_f} \left( 1-0.65\, n_f\frac{\alpha}{\pi}\right) \, .
    \end{cases} 
\label{BR_power_NLO_ortho}
\end{equation}

Figure~\ref{fig:comparison_photo_diss_NLO} compares the exact result with the corresponding approximations for $\mathcal{B}^{\rmi{1S},\rmi{para}}$ and $\mathcal{B}^{\rmi{1S},\rmi{ortho}}$, shown in the left and right panels, respectively. The smooth interpolation given in eq.~\eqref{approximate_diss_width_NLO} provides an excellent description at small $x$ and deviates from the exact result by at most about $15\%$ around $x\simeq2$. In contrast to the LO result, the NLO correction induces a mild dependence on the coupling $\alpha$, illustrated by the magenta and orange curves corresponding to $\alpha=0.03$ and $\alpha=0.1$, respectively. This behavior is observed for both para- and orthodarkonium.
The broken power-law approximation exhibits a qualitatively different behavior from its LO counterpart. While the LO broken power-law approximation systematically overestimates the exact rate, at NLO it instead underestimates the exact result at small $x$ and overestimates it at large $x$.
\subsection{Excited states}
\label{sec:excited_states}

So far we have focused on the ground state, whose contribution already captures the dominant qualitative features of bound-state effects. However, in the parameter range considered, excited states can provide non-negligible corrections to the effective annihilation cross section. This is particularly relevant when transitions among bound states are included \cite{Binder:2021vfo,Garny:2021qsr,Binder:2023ckj}. 

The treatment of excited states follows closely the discussion presented for the ground state. As an illustrative example, we consider the leading-order photo-dissociation width of the $2\mathrm{S}$ state. The exact expressions for the corresponding dissociation widths, together with those of the $2\mathrm{P}$ states and the bound-to-bound transition rates used in our numerical analysis, are collected in appendix~\ref{app:excited}.

The ionization cross section of the $2\mathrm{S}$ state reads~\cite{Biondini:2023zcz,Bethe:1957ncq}
\begin{align}
\sigma^{2\mathrm{S}}_{\rm ion}(|\bm{k}_\gamma|)
=
\alpha
\frac{2^{12}\pi^2}{3}\frac{|E_2|^3}{M |\bm{k}_\gamma|^4}
\frac{(4+w_2^2(|\bm{k}_\gamma|))}
{1+w_2^2}
\frac{
e^{-\frac{8}{w_2(|\bm{k}_\gamma|)}\arctan(w_2(|\bm{k}_\gamma|))}
}
{1-e^{-\frac{4\pi}{w_2(|\bm{k}_\gamma|)}}},
\label{ion_2S_exact}
\end{align}
where
\begin{align}
w_2(|\bm{k}_\gamma|)
=
\sqrt{\frac{|\bm{k}_\gamma|}{|E_2|}-1}.
\end{align}
Proceeding as for the ground state, the dissociation width can be written in terms of the dimensionless ratio
$x=|E_1|/T$. Since the first excited-state binding energy satisfies
$|E_2|=|E_1|/4$, one finds
\begin{align}
\Gamma^{2\mathrm{S}}_{\rm bsd}
=
\frac{2^2}{3}\,
\alpha^3 T \, 
x^3 \, 
\mathcal{F}^{2\mathrm{S}}(x),
\label{2S_exact_diss}
\end{align}
where the dimensionless function $\mathcal{F}^{2\mathrm{S}}(x)$ is defined in appendix~\ref{app:excited}, and the lower boundary of the thermal integral becomes $x/4$. 

Similarly to the ground-state case, the cross section is peaked close to threshold, implying that the thermal average is dominated by photon energies satisfying
$|\bm{k}_\gamma|\lesssim {\rm few}\times |E_2|$.
Unlike the $1\mathrm{S}$ state, however, the momentum dependence is modified by the additional polynomial factor, see~eqs.~\eqref{ion_1S_exact} and \eqref{ion_2S_exact}. It is therefore convenient to approximate the complete momentum dependence entering the thermal average. 
Expanding around threshold one finds up to $\mathcal{O}(w_2^4)$, the following expression 
\begin{align}
   \frac{1}{1+w_2^2} (4+w_2^2) \frac{
e^{-8\arctan(w_2)/w_2}
}
{1-e^{-4\pi/w_2}}\simeq \frac{1}{e^8} \left( 1+ \frac{8}{3} w_2^2\right) (4 - 3 w_2^2) = \frac{1}{e^8} \left( \frac{23}{3} \frac{|\bm{k}_\gamma|}{E_2} - \frac{11}{3}\right) 
\label{eq:BetheApprox2S}
\end{align}
where we used $w_2^2 = |\bm{k}_\gamma|/ E_2 -1$.
This  approximation is analogous to the one used for the 1S state, cf.~\eqref{eq:Bethe_approximation}, for the $2\mathrm{S}$.

\begin{figure}[t!]
    \centering
        \includegraphics[width=0.495\linewidth]{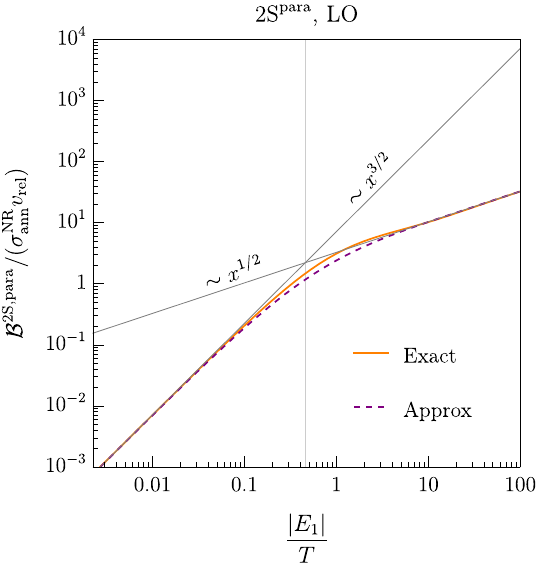}
        \includegraphics[width=0.48\linewidth]{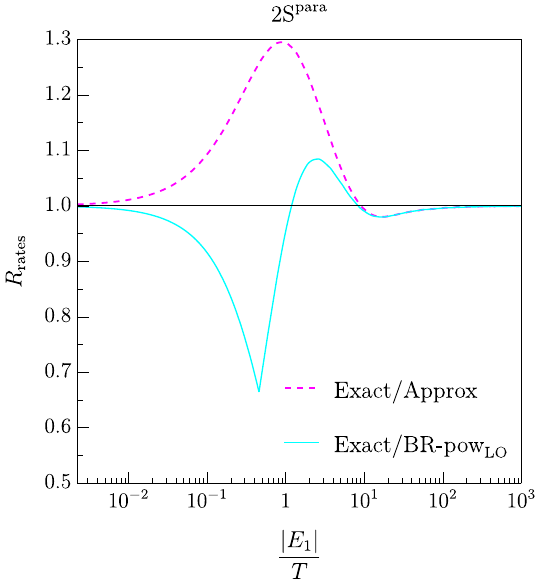}
    \caption{(Left) Effective rate capturing the 2S bound-state contribution to dark matter annihilation, normalized to the hard cross section $\sigma^{\hbox{\tiny NR}}_{\hbox{\scriptsize ann}} v_{\hbox{\scriptsize rel}}=\pi \alpha^2/\mDM^2$ for $n_f=0$. The orange line corresponds to the full pNRQED result, while the purple dashed line is obtained using the approximation in eq.~\eqref{2S_approximated_smooth}. The asymptotic behaviors at large and small $x$ are also indicated, which correspond to the broken power law eq.~\eqref{BR_power_LO_2S}. (Right) We show the ratio of the bound-state contribution in eq.~\eqref{def_effective_BS_rate} for the 2S state, computed using the dissociation width in eq.~\eqref{2S_exact_diss}, to the results obtained with the approximate width in eq.~\eqref{2S_approximated_smooth} and the power-law expression in eq.~\eqref{BR_power_LO_2S}, represented by the magenta-dashed and cyan solid lines.
    }
    \label{fig:comparison_photo_diss_2S}
\end{figure}

The asymptotic behaviour can again be understood analytically. In the low-temperature limit,
$x\gg 4$, the Bose-Einstein distribution reduces to
$1/(e^y-1)\simeq e^{-y}$, and the thermal integral is dominated by photon energies close to the dissociation threshold, $y\simeq x/4$. Expanding the integrand around threshold and employing the approximation in eq.~(\ref{eq:BetheApprox2S}), the explicit powers of $x$ cancel against the phase-space factor, leaving only the Boltzmann suppression associated with the binding energy of the $2\mathrm{S}$ state. Therefore,
\begin{align}
\Gamma^{2\mathrm{S}}_{\rm bsd}
\simeq 
\frac{1024}{3 \, e^8} \alpha^3 T
e^{-x/4},
\qquad
x\gg4.
\label{2S_bsd_xlarge}
\end{align}
Here the $1/4$ in the exponential arises from our choice of normalizing $T$ with respect to $E_1$ and not $E_2$ in the definition of $x$.
In the opposite limit,
$x\ll4$, the Bose-Einstein distribution behaves as
$1/(e^y-1)\simeq1/y$, such that the dissociation width scales linearly with the temperature,
\begin{align}
 \Gamma^{2\mathrm{S}}_{\rm bsd} \simeq  \frac{4480}{9 \, e^8} \alpha^3 T \qquad
x\ll4.
\label{2S_bsd_xsmall}
\end{align}

Exploiting the asymptotic behaviors, one may construct an interpolating expression for the $2\mathrm{S}$ dissociation width following the same strategy adopted for the ground state. An interpolating function that smoothly connects the limiting behaviors is 
    \begin{align}
\mathcal{B}^{\rmii{2S},\rmi{para}}_{\rmii{approx}}
=
\frac{\pi \alpha^2}{m_{\rm DM}^2}\,
\frac{2^4 \sqrt{\pi}\,2^{10}}{3e^8}\,
\frac{x^{1/2}}
{1+\dfrac{2^{12}}{3e^8}\dfrac{e^{-x/4}}{x}} \, .
\label{2S_approximated_smooth}
\end{align}
while the simple broken power law ansatz reads
\begin{equation}
\mathcal{B}^{\rmii{2S},\rmi{para}}_{\rmii{BR\,pow},\rmii{LO}} =
\begin{cases}
      2^2 \sqrt{\pi}\,
      \dfrac{\pi \alpha^2}{m_{\mDM}^2}\,
      x^{\frac{3}{2}} \,,
      & \quad
      x \leq \dfrac{2^{12}}{3e^8}\,,
      \\[4 mm]
      2^2 \sqrt{\pi}\,
      \dfrac{\pi \alpha^2}{m_{\rm DM}^2}\,
      \dfrac{2^{12}}{3e^8}\,
      x^{\frac{1}{2}} \,,
      & \quad
      x > \dfrac{2^{12}}{3e^8}\,.
\end{cases}
\label{BR_power_LO_2S}
\end{equation}

We show the effective rate capturing the $2S$ bound-state contribution to dark matter annihilation, normalized to the hard cross section, in figure~\ref{fig:comparison_photo_diss_2S}. The results can be compared with the corresponding ground-state contribution shown in figure~\ref{fig:comparison_photo_diss}. For ease of comparison, we retain the variable $|E_1|/T$ to characterize the temperature evolution, even though the binding energy of the $2S$ state is smaller by a factor of four. We observe a more pronounced difference between the smooth approximation and the broken power-law approximation in the 2S case. In particular, the approximate rate in eq.~\eqref{2S_approximated_smooth} underestimates the exact rate at large temperatures, with the discrepancy persisting down to $|E_1|/T \approx 8$. By contrast, the broken power-law approximation overestimates the exact rate in the high-temperature regime.
As discussed above, the contribution of a given bound state is peaked around its corresponding dissociation threshold. For the $2S$ state, this occurs at $|E_1|/T \approx 4$. Around this region, both the smooth approximation and the broken power-law approximation underestimate the exact rate, with a discrepancy of about 10\%.

The strategy presented in this section can also be applied to other excited
states, provided an appropriate approximation for the cross section close to
threshold is available. For the $2\mathrm{P}$ states, however, the treatment
differs from the $2\mathrm{S}$ case, at least at the order considered in this
paper. In particular, the annihilation width of the $2\mathrm{P}$ states
vanishes at leading order in the $1/\mDM$ expansion. Consequently, it is not
useful to construct a reduced rate $\mathcal{B}^{2\mathrm{P}_m}$ analogous to
the one introduced for the $2\mathrm{S}$ state, since such a contribution
would vanish at the order considered here. The $2\mathrm{P}$ dissociation
widths are nevertheless relevant when bound-to-bound transitions are included,
as they enter the corresponding transition matrix \cite{Garny:2021qsr,Binder:2021vfo}. We therefore approximate
the dissociation widths of the $2\mathrm{P}$ states directly, rather than
constructing a reduced rate.

The bound-state dissociation rates of the $2\mathrm{P}$ states can be computed
analogously to the $2\mathrm{S}$ case. Treating the magnetic quantum numbers
$m=0$ and $m=\pm1$ separately, we find in the low-temperature regime, $x \gg 4$, that the
$2\mathrm{P}_{m=0}$ dissociation rate is related to the $2\mathrm{S}$ rate in eq.~\eqref{2S_bsd_xlarge} by
a factor $49/36$, while the contribution from the two $m=\pm1$ states is
related by a factor $50/36$. Here, the latter result denotes the sum over the
two $m=\pm1$ states. The total $2\mathrm{P}$ dissociation rate is therefore
$11/4$ times the corresponding $2\mathrm{S}$ rate, in agreement with the ratio
of the corresponding photo-dissociation cross sections evaluated at threshold.
In the opposite, high-temperature regime, the $m=0$ and combined $m=\pm1$
dissociation rates approach constant coefficients multiplying $\alpha^3 T$.\footnote{With rational numbers, the corresponding expressions are
$\Gamma^{2\mathrm{P}_{m=0}}_{\rm bsd}
\simeq \frac{422912}{1225 e^8}\alpha^3 T$
and
$\Gamma^{2\mathrm{P}_{m=\pm1}}_{\rm bsd}
\simeq \frac{94208}{243 e^8}\alpha^3 T$.}
The corresponding expressions are given in appendix~\ref{app:excited}.


\section{Dark matter energy density}
\label{sec:numerical}
In this section we finally use the approximate rates and perform a quantitative study of the corresponding prediction for the energy density. Our aim is to offer alternative approaches to tackle the inclusion of non-perturbative dynamics in dark matter models. In section~\ref{sec:strategy_one} we solve the Boltzmann equation with the exact rates, as well as with their approximate versions as derived in section~\ref{sec:approx_bound_state_diss}. In the section~\ref{sec:strategy_two}, we instead take another viewpoint and first compare how well an \emph{approximate integral version} of the Boltzmann equation captures the non-perturbative dynamics. In the end, also in the second approach, we will use the approximate rates and assess the effect of combining these simplifications. 
\begin{figure}[t!]
    \centering
    \includegraphics[width=0.5\linewidth]{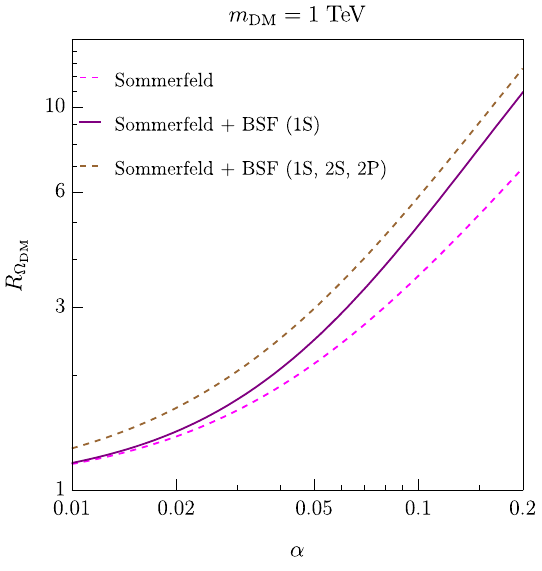}
    \caption{Ratio of the relic density obtained in the pure $s$-wave case to that obtained by successively including Sommerfeld enhancement (magenta dashed), ground-state bound-state formation (purple-solid), and excited bound states with transitions among them (brown-dashed), as a function of the coupling $\alpha$.
}
    \label{fig:example_effects}
\end{figure}

To illustrate the impact of near-threshold effects on the DM relic density, and before assessing the errors introduced by our approximate rates, we compare the relic density predictions obtained with different levels of approximation. In figure~\ref{fig:example_effects}, we show the ratio of the relic density obtained in the pure $s$-wave case to that obtained when including near-threshold effects. We consider, in turn, the inclusion of Sommerfeld enhancement, the additional contribution of ground-state bound-state formation, and finally the inclusion of excited states and transitions among them. We find that neglecting near-threshold effects can lead to an underestimation of the relic density by up to about one order of magnitude for couplings $\alpha \gtrsim 0.1$.

    

\subsection{Numerical solution of the Boltzmann equation}
\label{sec:strategy_one}
In this section we numerically solve the Boltzmann equation outlined in eqs.~\eqref{Boltzmann_eq_eff} and \eqref{Cross_section_eff}. The main goal is to assess how accurately the approximate rates reproduce the DM relic abundance compared to the exact rates. One may already anticipate the outcome from the plots in the right panel of figure~\ref{fig:comparison_photo_diss} and in figure~\ref{fig:comparison_photo_diss_NLO}. However, a quantitative assessment requires solving the Boltzmann equation numerically and evaluating how the differences in the rates propagate to the DM yield. Indeed, the discrepancy among the various rates is non-trivially weighted during the thermal evolution and, as already mentioned, the bound-state effects are important close to the corresponding energy threshold $T \simeq |E_n|$. 
For the sake of clarity, we first focus on the error coming from approximating the bound-state rates while keeping the Sommerfeld factor exact. In a second stage, we then investigate the impact of approximating the Sommerfeld factor as well. 

As customary in the numerical solution of the Boltzmann equation, we introduce the DM yield as $Y=n/s$, where $s$ is the entropy density. In a radiation-dominated Universe, the latter is given by
$s=\frac{2\pi^2}{45}T^3 h_{\rmi{eff}}(T)$,
where $h_{\rmi{eff}}$ are the effective entropic degrees of freedom. In our numerical study we take these as well as the effective number of relativistic degrees of freedom $g_{\rmi{eff}}$, which enter the Hubble rate, from ref.~\cite{Laine:2015kra}. In terms of the yield, the dilution due to the expansion of the Universe is factored out, since both the number and entropy densities scale as $a^{-3}$ in the absence of entropy production and number-changing processes. The Boltzmann equation~\eqref{Boltzmann_eq_eff} then becomes
\begin{eqnarray}
\frac{dY}{dz}
=-\frac{1}{2}\frac{s \xi(T)}{zH}
\langle \sigma_{\rmi{eff}} v_{\textrm{rel}}\rangle
\left(Y^2-Y_{\textrm{eq}}^2\right) \, ,
\label{BE_version_yield}
\end{eqnarray}
where $z=\mDM/T$ is a convenient dimensionless time variable for following the evolution of the yield, and the temperature-dependent function $\xi(T)$ is defined as
\begin{equation}
\xi(T)\equiv
1+\frac{d\ln h_{\textrm{eff}}(T)}{3 \, d\ln T} \, .
\end{equation}
The present-day DM energy density can then be expressed in terms of the asymptotic yield as
\begin{equation}
\Omega_{\rmii{DM}}h^2
=\frac{\mDM}{\text{GeV}}
\frac{Y(z_{\rmi{fin}})}{3.645\times10^{-9}} \, ,
\label{DM_energy_density_from_Y}
\end{equation}
where $Y_Y(z_{\rmi{fin}})$ denotes the DM yield at a reference temperature $T_{\rmi{fin}}$ below which the annihilation processes no longer modify the yield appreciably. In our numerical analysis, we choose this temperature such that the residual change in the yield is below 1\%, corresponding to the accuracy required for the comparison with the experimental determination of the relic abundance. The numerical coefficient in eq.~\eqref{DM_energy_density_from_Y} follows from the present-day entropy and critical energy densities taken from ref.~\cite{ParticleDataGroup:2016lqr}.

\begin{figure}[t!]
    \centering
    \includegraphics[width=0.5\linewidth]{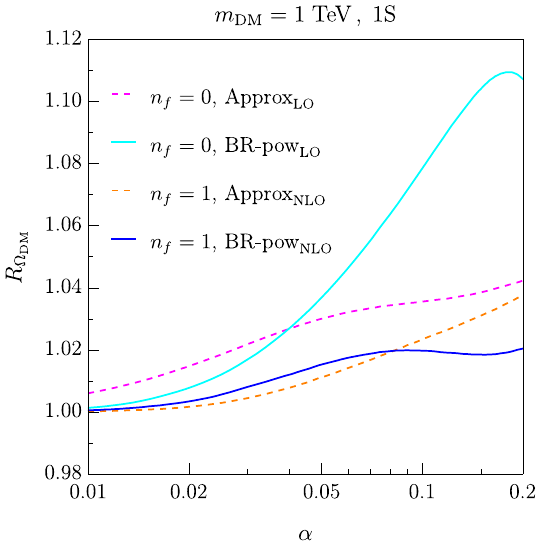}
    \caption{Ratio of the relic density as a function of $\alpha$ obtained using the exact rates to that obtained using the corresponding approximate rates. Only the ground state is included. The smooth approximations are shown by the magenta-dashed and orange-dashed lines for the LO and NLO rates, respectively, while the cyan-solid and blue-solid lines correspond to the respective broken power-law approximations. 
    }
    \label{fig:comparison_relic_density_num}
\end{figure}

When only the ground state is included, the discrepancy induced by the approximate rates is shown in figure~\ref{fig:comparison_relic_density_num}. We display the ratio between the relic density obtained using the exact rates and the corresponding result obtained with the approximate expressions. This parallels the discussion outlined in section~\ref{sec:approx_bound_state_diss}, so that one may recognize the same trend from the rates to the DM energy density.
The magenta dashed and cyan solid curves correspond to the case $n_f=0$. In this scenario, only the paradarkonium contributes to annihilation through bound states. Accordingly, NLO corrections are absent, since they vanish for $n_f=0$ ( cf.~eqs~\eqref{bsf_final_fixed_NLO}and \eqref{bsd_final_fixed_NLO}). This setup therefore corresponds to including only the leading-order processes, namely bound-state formation and dissociation via dark  photon emission and absorption respectively.

As shown in figure~\ref{fig:comparison_photo_diss} (left), the broken power-law approximation for the bound-state contribution, $\mathcal{B}^{\rmii{1S},\rmi{para}}_{\rmi{BR\,pow},\rmii{LO}}$, systematically overestimates the exact rate, and does so more significantly than $\mathcal{B}^{\rmii{1S},\rmi{para}}_{\rmi{approx},\rmii{LO}}$. This translates into the largest deviation in the predicted relic density, represented by the cyan solid curve, reaching about $11\%$ for the largest values of $\alpha$. The discrepancy is already substantially reduced when the approximate rate in eq.~\eqref{effective_rate_para_approx} is used to account for bound-state effects. In the absence of light fermions, the error in the relic density prediction remains below $4\%$ even at $\alpha=0.2$.

Let us now turn to the NLO rates. The smooth interpolations in eqs.~\eqref{approximate_diss_width_NLO} and \eqref{barX2_averaged_NLO} exhibit a qualitatively similar behavior to its LO counterpart. In particular, it systematically overestimates the exact rate (see figure~\ref{fig:comparison_photo_diss_NLO}), implying that the dark matter relic abundance obtained using the approximate NLO rate is always slightly smaller than that obtained from the exact NLO result. The discrepancy depends only mildly on the coupling: the orange and magenta curves display a comparable level of accuracy, with deviations remaining below approximately 4\% over the entire range of couplings considered.

The situation is qualitatively different for the broken power-law approximation. Unlike at LO, where the approximate rate consistently overestimates the exact result, the inclusion of the NLO corrections causes the broken power-law approximation to both underestimate and overestimate the exact rate, depending on the value of $x$. The size of this effect is controlled by the coupling. For larger values of $\alpha$, the approximate rate significantly underestimates the exact result at small $x$, before crossing over to an overestimation at larger $x$. As $\alpha$ decreases, the low-$x$ underestimation becomes progressively weaker, whereas the overestimation at large $x$ remains essentially unchanged. Consequently, the cancellation between the low- and high-$x$ errors depends sensitively on the coupling, leading to a non-trivial $\alpha$ dependence in the predicted relic abundance.

Despite this more intricate behavior, the broken power-law approximation ultimately provides a remarkably accurate prediction of the relic abundance, with a precision comparable to that obtained using the smooth interpolation of the NLO rate. This suggests that the simpler broken power-law approximation, constructed \emph{solely} from the asymptotic small- and large-temperature limits, already captures the essential physics governing bound-state effects and is sufficient to predict the relic density at the level of about $2\%$.

\begin{figure}[t!]
    \centering
    \includegraphics[width=0.5\linewidth]{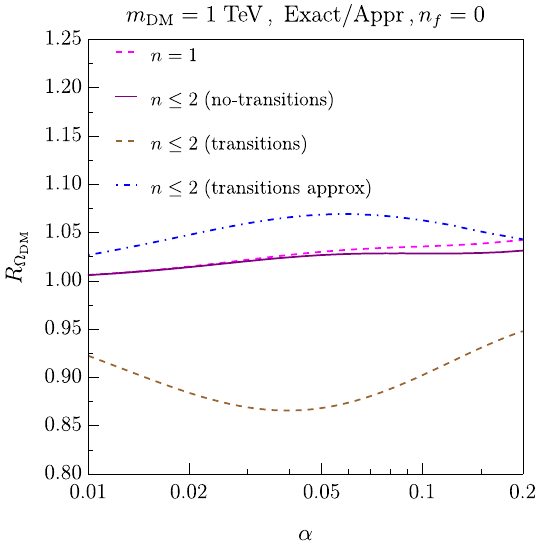}
    \caption{This plot shows the ratio of the relic density obtained using different treatments of the bound-state contribution. The magenta curve includes only the $1S$ bound state. The purple curve additionally includes the contribution from the excited $2S$ state, while neglecting bound-to-bound transitions. The brown dashed line shows the ratio between the relic density obtained using the exact rates with five bound states, including transitions among them, and the result obtained with the approximated rates including only the $1S$ and $2S$ states in the no-transition limit. The blue dot-dashed line takes as an input for the transitions the approximated dissociation widths.}
    \label{fig:relic_density_ratio_excited}
\end{figure}

Finally, we turn to the excited states and we aim to show the overall error introduced by our approximations, even including the approximate thermally averaged Sommerfeld factor in eq.~\eqref{approx_Sommerfeld}. As discussed in Section~\ref{sec:excited_states}, one may also construct an approximation for the reduced width entering $\mathcal{B}^n$, cf.~\eqref{def_effective_BS_rate}, for the $2S$ state. We then assess the impact of the smooth approximation on the extraction of the DM relic density when the $2S$ state is included. Our results are shown in Fig.~\ref{fig:relic_density_ratio_excited}. The dashed-magenta curve is not the same as that shown in Fig.~\ref{fig:comparison_relic_density_num}, since it now contains the effect of the approximate Sommerfeld factor. 
The solid-purple curve shows the ratio between the relic density obtained using the exact bound-state dissociation rates for the para-darkonium $1S$ and $2S$ states and that obtained using the corresponding smooth approximations, eqs.~\eqref{effective_rate_para_approx} and~\eqref{2S_approximated_smooth}, respectively. Here, transitions among bound states are neglected. The discrepancy remains below approximately $3.0\%$ over the range of $\alpha$ considered, reaching its largest value at the largest couplings.

As a further comparison, we consider the ratio between the relic density obtained with bound-state transitions included and using the exact bound-state dissociation rates, and that obtained in the no-transition limit, including both the $1S$ and $2S$ states and using the smooth approximations for their dissociation rates. This corresponds to the dashed-brown curve in Fig.~\ref{fig:relic_density_ratio_excited}. The effect of bound-state transitions is found to be more prominent, as also emphasized in earlier studies, leading to deviations of up to approximately $15\%$ in the relic density ratio. Finally, the blue dot-dashed line includes the transitions, with all dissociation rates obtained using our approximations. As one can see, the agreement improves once the transitions are included, reducing the discrepancy by roughly a factor of two, even when the rates are treated approximately. It is worth remarking, however, that this does not alleviate the complexity of deriving the transition matrix elements.

At this stage, however, we cannot further simplify the treatment of the excited states by resorting to the so-called \emph{efficient transition limit}~\cite{Garny:2021qsr}. In this limit, the transition rates must be sufficiently large compared with both the bound-state decay and photodissociation rates, for both directions of the relevant transitions. In particular, for the $2P_m\leftrightarrow1S$ transitions, this requires
\begin{align}
    \Gamma^{\rmi{2P}_m \to \rmi{1S}}
    &\gg
    \Gamma_{\rmi{dec}}^{\rmi{2P}_m},
    \Gamma_{\rmi{bsd}}^{\rmi{2P}_m},
    \qquad
    \Gamma^{\rmi{1S} \to \rmi{2P}_m}
    \gg
    \Gamma_{\rmi{dec}}^{\rmi{1S}},
    \Gamma_{\rmi{bsd}}^{\rmi{1S}} .
\end{align}
We have explicitly checked these hierarchies and find that they are not simultaneously satisfied for bound-state decay and bound-state photodissociation over the parameter range considered. Therefore, the efficient-transition limit cannot be consistently applied to our system.

\subsection{Semi-analytical solution of the Boltzmann equation}
\label{sec:strategy_two}

In practice, one might be interested in avoiding the full numerical
solution of the Boltzmann equation in order to simplify the calculation
and reduce the computational effort. In this section we show that this is indeed possible while retaining a very good agreement with the full numerical solution, \emph{and} accounting for near-threshold effects.

As discussed above, Sommerfeld
enhancement and bound-state effects become more important at lower
temperatures, whereas their impact at high temperatures is relatively
small. Therefore, we expect the initial departure from thermal
equilibrium to proceed in a similar way with and without near-threshold
effects, while the latter become more important during the subsequent
evolution.

This motivates a semi-analytical treatment of the Boltzmann equation,
following the standard freeze-out approximation \cite{Kolb:1990vq}, see also \cite{vonHarling:2014kha} for
an early application in the context of bound states. We summarize the
main steps here. Once the dark matter yield has started to depart from
chemical equilibrium, the ratio $Y/Y_{\rm eq}$ increases rapidly due to
the exponential suppression of $Y_{\rm eq}$. In this regime, the
back-reaction can be neglected and eq.~\eqref{BE_version_yield} becomes
\begin{equation}
    \frac{dY}{dz}
    =
    -\frac{1}{2}
    \frac{s\xi(T)}{zH}
    \langle\sigma_{\rmi{eff}} v_{\rm rel}\rangle Y^2 .
    \label{eq:boltzmann_semi}
\end{equation}
This equation can be solved by direct integration and gives
\begin{equation}
    \frac{1}{Y(z_{\rm fin})}
    -
    \frac{1}{Y(z_{\rm fo})}
    =
    \int_{z_{\rm fo}}^{z_{\rm fin}} dz\,
    \frac{1}{2}
    \frac{s\xi(T)}{zH}
    \langle\sigma_{\rmi{eff}} v_{\rm rel}\rangle ,
    \label{eq:boltzmann_integral}
\end{equation}
where $z_{\rm fo}$ denotes the freeze-out temperature. Defining the
latter by
\begin{equation}
    \frac{Y(z_{\rm fo})-Y_{\rm eq}(z_{\rm fo})}
    {Y_{\rm eq}(z_{\rm fo})}
    = c ,
    \label{eq:freezeout_condition}
\end{equation}
where $c$ is an order-one parameter, the yield at the freeze-out
temperature entering eq.~(\ref{eq:boltzmann_integral}) can be written as $Y(z_{\rm fo})=(1+c)Y_{\rm eq}(z_{\rm fo})$.

We note in passing that the integral on the right-hand side of eq.~\eqref{eq:boltzmann_integral} can in some cases also be evaluated analytically. If the
variation of the effective number of degrees of freedom is neglected
during the relevant part of the evolution, the approximate Sommerfeld
factor in eq.~\eqref{approx_Sommerfeld} and the broken power-law approximations for the
bound-state contribution in eqs.~\eqref{BR_power_LO_para}, \eqref{BR_power_LO_ortho} lead to analytic
expressions for the integral.  In the following, we do not use these two simplifications and we then numerically evaluate the integral appearing in eq.~(\ref{eq:boltzmann_integral}) in all cases. 


The remaining ingredient is the determination of $z_{\rm fo}$. A precise
determination of this quantity would require solving the Boltzmann
equation numerically. However, the final result depends only weakly on
its value, and the effects of Sommerfeld
enhancement and bound-state formation are still  small at the relatively
high temperatures relevant for the onset of freeze-out. We therefore
expect that an approximate determination of $z_{\rm fo}$ is sufficient
for the purpose of evaluating eq.~(\ref{eq:boltzmann_integral}).

In the absence of a prior numerical solution, we determine $z_{\rm fo}$
using the standard analytical freeze-out approximation \cite{Kolb:1990vq}. Defining
the departure from chemical equilibrium as
\begin{equation}
    \Delta = Y-Y_{\rm eq}\
\end{equation}
one can recast the Boltzmann equation in terms of $\Delta$. After neglecting
$d\Delta/dz$ compared to $\Delta$, $Y_{\rm eq}$ and $dY_{\rm eq}/dz$,
which is justified as long as $Y_{\rm eq}\sim\Delta$, this leads to an
algebraic relation between $\Delta$ and $z$. Keeping \emph{only} the $s$-wave
contribution, i.e.~without near-threshold effects, this relation can be solved for the freeze-out
temperature, giving
\begin{equation}
    z_{\rm fo}
    =
    \frac{1}{2}
    W_0\left[
        \frac{\big(a\lambda c(2+c)\big)^2}{2}
    \right],
    \label{eq:zfo_semi}
\end{equation}
where $W_0$ denotes the zero branch of the Lambert $W$ function. The
quantities $a$ and $\lambda$ are given by
\begin{equation}
    a =
    \frac{45}{4\sqrt{2}\,\pi^{7/2}}
    \frac{g_X}{h_{\rm eff}},
    \qquad
    \lambda =
    \left.
    \frac{s\langle\sigma_{\rmi{eff}} v_{\rm rel}\rangle}{H}
    \right|_{T=m_{\rm DM}}
    =
    \sqrt{\frac{\pi}{45}}
    \frac{h_{\rm eff}}{\sqrt{g_{\rm eff}}}
    M_{\rm Pl} \mDM
    \langle\sigma_{\rmi{eff}} v_{\rm rel}\rangle .
    \label{eq:lambda}
\end{equation}
In contrast to the treatment in ref.~\cite{Kolb:1990vq}, we do not expand the Lambert
$W$ function. We also note that the factor $1/2$ in the argument of the
Lambert function follows from the normalization of the Boltzmann
equation used here.

The value of $c$ in eq.~(\ref{eq:freezeout_condition}) is not fixed by
the physics and can be chosen to optimize the approximation. A
conventional choice is $c=\sqrt{2}$ \cite{Kolb:1990vq}. In existing
automated tools for the computation of the dark matter relic density,
such as DarkSUSY \cite{Gondolo:2004sc}, micrOMEGAs \cite{,Belanger:2006is} and MadDM
\cite{Backovic:2013dpa}, freeze-out is
also determined using an order-one criterion of this type; in micrOMEGAs one finds $c=1.5$. We
compared the choice $c=\sqrt{2}$ with $c=1.5$ and find that the
difference in the freeze-out temperature obtained from the numerical
and semi-analytical determinations remains below $0.2\%$ over the range
of $\alpha$ considered in this work. The choice $c=1.5$ gives a
slightly smaller difference with the numerically-extracted $z_{\rm{fo}}$ and is therefore used in the following. 

It is important to stress that eq.~(\ref{eq:zfo_semi}) is used only to
determine the starting point of the post-freeze-out evolution. The
Sommerfeld and bound-state effects are not included in this expression.
Instead, they are included in the subsequent evolution through the
cross section entering the integral in
eq.~(\ref{eq:boltzmann_integral}). In this way, the semi-analytical
procedure separates the determination of the freeze-out point from the
effects that become increasingly important at later times.

As an alternative to eq.~(\ref{eq:zfo_semi}), one can use a freeze-out
temperature obtained from an existing numerical freeze-out calculation.
For example, public dark matter codes provide a value of $z_{\rm fo}$
according to a freeze-out criterion of the form in
eq.~(\ref{eq:freezeout_condition}). This value can simply be used as an
input to eq.~(\ref{eq:boltzmann_integral}), without requiring a second
numerical solution of the post-freeze-out evolution. This provides a
simple way of refining an otherwise standard freeze-out calculation by
including the Sommerfeld and bound-state effects in the subsequent
evolution.

\begin{figure}[t!]
    \centering
    \includegraphics[width=0.48\linewidth]{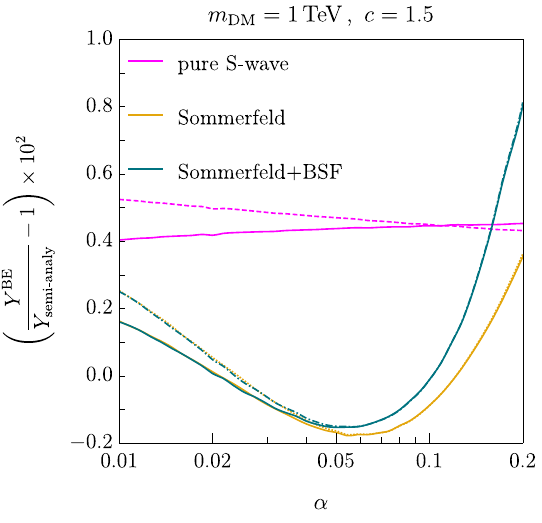}
    \includegraphics[width=0.45\linewidth]{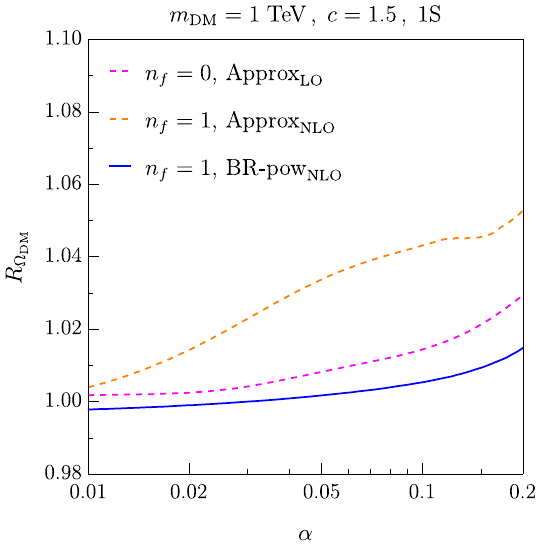}
    \caption{(Left) Relative difference in the yield, evaluated at $z_{\textrm{fin}} = 10^5$, expressed as a percentage and plotted as a function of $\alpha$. To extract $Y_{\textrm{semi-analy}}$, we use two different inputs for the freeze-out variable $z_{\textrm{fo}}$: (i) the value obtained from eq.~\eqref{eq:zfo_semi} (solid lines), and (ii) the value obtained by solving the Boltzmann equation \eqref{Boltzmann_eq_eff} and imposing $(Y - Y_{\textrm{eq}})/Y_{\textrm{eq}} = c$ (dashed lines). The magenta, gold and green lines stand respectively for the pure $s$-wave, Sommerfeld enhanced, and Sommerfeld and bound-state effects. Only the ground state is included. The relative difference remains below $1\%$ in all cases. (Right) Ratio of the relic density obtained from the numerical solution of the Boltzmann equation with exact rates to that obtained using the semi-analytical approach to the rate equation and approximate rates.}
    
    \label{fig:zfo_comparison}
\end{figure}

We now compare the final yield obtained from the semi-analytical
prescription with the numerical solution of the Boltzmann equation. We
consider two procedures for determining the final yield:
\begin{enumerate}
    \item solving the Boltzmann equation numerically, which we denote
    by the superscript ``BE'';
    \item using the semi-analytical prescription, namely determining
    $z_{\rm fo}$ from eq.~(\ref{eq:zfo_semi}) and evaluating the integral
    in eq.~(\ref{eq:boltzmann_integral}), which we denote by the
    superscript ``semi-analy''.
\end{enumerate}
Note that, to isolate the effect of the semi-analytical approximation, we use the exact rates here. The extension to approximate rates is discussed further below.
The comparison is shown in figure~\ref{fig:zfo_comparison}, right panel,
where the relative difference between the two yields is displayed at
$z_{\rm fin}=10^5$. For completeness, we also show the result obtained
when the numerically determined value of $z_{\rm fo}$ is used in the integral solution \eqref{eq:boltzmann_integral}. The magenta, gold and green curves correspond
to the pure $s$-wave, Sommerfeld-enhanced, and Sommerfeld-enhanced plus
bound-state cases, respectively. Only the ground state is included.
The relative difference between the semi-analytical and numerical
solutions remains below $1\%$ in all cases. As expected, the impact of bound states is negligible at small $\alpha$, where the green and gold lines overlap. At large $\alpha$, the yield is instead controlled by late-time annihilations, making the choice of $z_{\rm fo}$ largely irrelevant.

These results show that, for the ground-state contribution considered
here, the post-freeze-out evolution can be accurately described by the
integral in eq.~(\ref{eq:boltzmann_integral}). In particular, the
Sommerfeld enhancement and bound-state contribution can be included
without solving the full Boltzmann equation again, once a suitable
freeze-out temperature is provided. The semi-analytical treatment
therefore gives a simple way of extending a standard freeze-out
calculation to include the near-threshold effects studied in this work.

Finally, in the right panel of figure~\ref{fig:zfo_comparison}, we show the ratio of the relic density obtained from the numerical solution of the Boltzmann equation with exact rates to that obtained using our semi-analytical approach with approximate rates. This comparison summarizes the error introduced by our framework, which combines physically motivated approximations for the rates with an integral equation rather than the full differential Boltzmann equation. We emphasize that, over the entire range of couplings considered, our approach, incorporating approximate Sommerfeld enhancement and ground-state bound-state effects, reproduces the relic density obtained using exact rates and a numerical solution of the Boltzmann equation to within $\sim5\%$.

The approximation should, however, be understood within the assumptions
used to derive the effective cross section. In particular, the discussion
above applies directly to the ground-state, no-transition case. As
shown in section~4.1, transitions among excited states can have a
more sizeable impact on the relic abundance, while the efficient-transition
limit is not consistently realized over the parameter range considered
here. In such cases, the coupled evolution of the bound-state
populations has to be treated explicitly if a precision of better than $15\%$ is required since  the simplification to a
single integral does not by itself remove the need for the corresponding
transition rates.

\section{Conclusions}
\label{sec:conclusions}

In this work, we have studied how Sommerfeld enhancement and
bound-state effects can be included in the computation of the dark
matter relic abundance in a simple and efficient way. We have focused
on an abelian dark matter model with a massless vector mediator and used
the potential non-relativistic effective field theory framework to
organize the relevant interaction rates. Our main goal was to identify
simple approximations that retain the relevant temperature dependence
of the rates while reducing the computational effort required for a
numerical relic-density calculation.

We first considered the Sommerfeld-enhanced annihilation rate and the
bound-state contribution to the effective cross section. By studying
the high- and low-temperature limits of the corresponding rates, we
constructed simple approximations that reproduce the main features of
the full results. In particular, we found that the Sommerfeld factor
can be described by a smooth interpolation between its two asymptotic
regimes with a small deviation from the full result (below 4\%). For the
bound-state contribution, we considered both smooth interpolating
functions and broken power-law approximations based on the analytic
limits of the dissociation rate. These approximations provide a simple
way of evaluating the bound-state contribution without performing the
full thermal integrations for every parameter point.

We then studied how these approximations affect the dark matter relic
abundance by solving the Boltzmann equation numerically. For the
ground-state contribution, we find that the smooth approximation gives
a good description of the relic abundance over the range of couplings
considered. The comparison with the exact rates shows that the
deviation remains at the few-percent level, while the broken power-law
approximation can lead to somewhat larger differences, in particular
at larger values of the coupling. Including the NLO corrections
improves the agreement of the broken power-law approximation with the
full result (maximal 2\% error). We have also checked the dependence on the dark matter
mass and find that the corresponding differences remain at the
percent level for the mass range considered.

We have also investigated the effect of excited bound states. Including the 2S state leads to a moderate change in the relic abundance, and we provide the corresponding approximate rate to account for it in a straightforward way. When transitions among several bound states are included, the relic abundance can deviate from the ground-state-only result by up to approximately $15\%$, as found in our numerical analysis.
 We have
checked the conditions required for the efficient-transition limit and
find that they are not simultaneously satisfied for bound-state decay
and photodissociation in our setup. Therefore, this limit cannot be
used consistently to further simplify the treatment of the excited
states. A treatment including such transitions still requires the
corresponding transition matrix and does not provide the same
simplification as in the ground-state, no-transition case. When bound-state transitions are included using the approximate rates
derived in this work, the resulting relic abundance differs from the
exact treatment by at most approximately $5\%$.

Finally, we considered a semi-analytical treatment of the Boltzmann equation. After freeze-out, the equilibrium contribution can be neglected, allowing the Boltzmann equation to be integrated directly and reducing the determination of the final yield to a single integral. The freeze-out temperature can be obtained either from the standard analytical freeze-out approximation, as in eq.~\eqref{eq:zfo_semi}, or from a numerical freeze-out calculation, for instance, using an automated tool. Using this value in the post-freeze-out integral leads to a final yield that agrees with the full numerical solution to better than $1\%$ for the pure $s$-wave, Sommerfeld-enhanced, and Sommerfeld-enhanced plus ground-state bound-state cases. When combining the semi-analytical approach with our approximate rates, we find that the resulting error remains below $5\%$ over the entire range of couplings considered in this work.

These results show that, in the regime considered here, Sommerfeld and
ground-state bound-state effects can be included in a standard
freeze-out calculation with relatively little additional
computational effort. 

The simple prescription discussed here is not intended to replace a
full numerical treatment in cases where excited-state transitions or
other effects invalidate the assumptions entering the effective
description and percent level accuracy is desired. Rather, it provides a complementary approach that can be
used to estimate the impact of Sommerfeld enhancement and bound-state
effects with limited effort.
We have focused here on an
abelian dark matter model, but we expect that our strategy can be
extended to  other models with significant near-threshold effects, e.g. models with non-abelian gauge symmetries.
In this case, the relevant interaction
rates are more involved, but approximations based on their analytic
high- and low-temperature limits may provide a similar way of
simplifying the relic-density calculation.
\appendix
\renewcommand{\thesection}{\Alph{section}}
\renewcommand{\thesubsection}{\Alph{section}.\arabic{subsection}}
\renewcommand{\theequation}{\Alph{section}.\arabic{equation}}

%
\section{Excited states}
\label{app:excited}
Here we collect the widths of the excited states that we have considered in the numerical section~\ref{sec:numerical}. We provide the explicit expression at LO, while the NLO ones can be easily inferred from the general formulas in eqs.~\eqref{bsd_final_fixed_NLO}. The 2S dissociation width reads
\begin{eqnarray}
    \Gamma^{2 \textrm{S}}_{\textrm{bsd}} &=&  \frac{2^2}{3} \alpha^3 \, T \left(  \frac{|E_1|}{T} \right)^3 \, 
   \mathcal{F}^{2 \textrm{S}}\left(  \frac{|E_1|}{T} \right)  \, ,
   \label{exact_diss_widths_2S}
\end{eqnarray}
where
\begin{eqnarray}
    \mathcal{F}^{2 \textrm{S}}_{\textrm{bsd}}(x) = \int_{\frac{x}{4}}^{\infty} d y \, \frac{(4+ w_2(x,y)^2)}{y^3 (e^y-1)} \frac{e^{-8 \frac{\arctan(w_2(x,y))}{w_2(x,y)}}}{1-e^{-4 \pi/w_2(x,y)}} \, ,
    \label{integral_diss_widths_2S}
\end{eqnarray}
where we have traded the appearance of $|E_2|$ with the ground state binding energy through $|E_2|/|E_1|=1/4$ and, therefore, and $w_2(x,y)=\sqrt{4y/x-1}$, where $y=|\bm{k}_\gamma|/T$ and $x=|E_1|/T$.

For the bound-state dissociation of 2P states we have to differentiate between the different projections of the angular momentum, as labeled with 2P$_0$ and 2P$_{\pm 1}$, and we obtain 
\begin{eqnarray}
    \Gamma^{2 \textrm{P}_{0}}_{\textrm{bsd}} &=&  \frac{1}{3} \alpha^3 \, T \left(  \frac{|E_1|}{T} \right)^4 \, 
   \mathcal{F}^{2 \textrm{P}_{0}}\left(  \frac{|E_1|}{T} \right)  \, ,
   \label{exact_diss_widths_2P0}
   \\
    \Gamma^{2 \textrm{P}_{\pm 1}}_{\textrm{bsd}} &=&  \frac{1}{3} \alpha^3 \, T \left(  \frac{|E_1|}{T} \right)^4 \, 
   \mathcal{F}^{2 \textrm{P}_{\pm 1}}\left(  \frac{|E_1|}{T} \right)
    \label{exact_diss_widths_2P1}
\end{eqnarray}
where
\begin{eqnarray}
    \mathcal{F}^{2 \textrm{P}_{0}}_{\textrm{bsd}}(x) &=& \int_{\frac{x}{4}}^{\infty} d y \, \frac{\left( \sqrt{w_2(x,y)^2+1} - 4 \sqrt{w_2(x,y)^2+4}\right)^2}{y^4 (e^y-1)} \frac{e^{-8 \frac{\arctan(w_2(x,y))}{w_2(x,y)}}}{1-e^{-4 \pi/w_2(x,y)}} \, ,
    \label{aux_diss_widths_2P0} 
    \\
     \mathcal{F}^{2 \textrm{P}_{\pm 1}}_{\textrm{bsd}}(x) &=& \int_{\frac{x}{4}}^{\infty} d y \, \frac{\left( \sqrt{w_2(x,y)^2+1} + 2 \sqrt{w_2(x,y)^2+4}\right)^2}{y^4 (e^y-1)} \frac{e^{-8 \frac{\arctan(w_2(x,y))}{w_2(x,y)}}}{1-e^{-4 \pi/w_2(x,y)}} \, .
    \label{aux_diss_widths_2P1} 
\end{eqnarray}

The final ingredients are the excitations and de-excitation widths. Their expression is quite simple and they read, also at LO, as follows  
\begin{eqnarray}
    \Gamma^{1\textrm{S}\rightarrow 2\textrm{P}}_{\textrm{ex.}} &=&
  \frac{2^7}{3^7}\frac{M\alpha^5}{e^{\frac{3M\alpha^2}{16T}}-1} \, , 
  \\
  [2pt]
   \Gamma^{2\textrm{P}\rightarrow 1\textrm{S}}_{\textrm{de-ex.}}
  &=&\frac{2^7}{3^8}\frac{M\alpha^5}{1-e^{-\frac{3M\alpha^2}{16T}}} \, .
\end{eqnarray}


\newpage
\bibliographystyle{JHEP2.bst}    
\bibliography{references}
\end{document}